\documentclass[aps,pra,twocolumn,preprintnumbers,superscriptaddress,nofootinbib,floatfix]{revtex4-2}

\usepackage{amsmath,amssymb}
\usepackage{braket}
\usepackage{bm}
\usepackage{graphicx}
\usepackage{tikz}
\usetikzlibrary{quantikz,arrows.meta}
\usepackage[caption=false]{subfig}

\newcommand{\Id}{\mathbb{I}}

\begin{document} \preprint{FERMILAB-PUB-26-0826-ETD}

\title{Quantum Zeno Effect with a Superposed Measurement Schedule in Relational Time}

\author{O. Seidel} \email{olivias@fnal.gov}
\affiliation{Fermilab Quantum Division, Batavia, Illinois, USA}
\affiliation{Department of Physics, University of Texas at Arlington, Arlington, Texas, USA}
\author{C. Marletto}
\affiliation{Department of Physics, University of Oxford, Oxford, United Kingdom}
\author{J. Asaadi}
\affiliation{Department of Physics, University of Texas at Arlington, Arlington, Texas, USA}
\author{S. Sussman}
\affiliation{Fermilab Quantum Division, Batavia, Illinois, USA}
\author{S. Biswas} \affiliation{Department of Physics, University of Texas at Arlington, Arlington, Texas, USA} \author{V. Vedral}
\affiliation{Department of Physics, University of Oxford, Oxford, United Kingdom}

\date{\today}

\begin{abstract}
Motivated by the Page--Wootters construction in which time is a quantum clock reading and the global system state is stationary, we propose a quantum circuit that explores the observable consequences of treating measurement timing as a quantum degree of freedom. We place two Zeno measurement schedules of a driven qubit in a coherent superposition using a schedule qubit. We show that, for a fixed drive strength, the interference visibility between the two schedules depends only on the time offset between them, not on the number of intermediate measurements taken. Then we show that, for one schedule-readout outcome, the Zeno trend reverses, predicting a result that complements the standard Zeno probability. We propose a circuit implementation with physical memory qubits and show that measuring the memory register's joint $X$-basis parity distinguishes a coherent schedule from a classical or dephased one. We give the gate count and an error budget for a superconducting-qubit implementation. The Page--Wootters picture motivates this protocol and provides an equivalent stationary representation, although the predictions follow from the standard circuit description without requiring a relational-time interpretation. This is consistent with the Page--Wootters framework recovering standard quantum dynamics through correlations between a clock and the rest of the system.
\end{abstract}

\maketitle

\section{Introduction}
Few effects in quantum mechanics depend as directly on when a measurement is made as the quantum Zeno effect. A driven qubit that is measured often enough stays close to its initial state, while the same system measured less often evolves away from it~\cite{MisraSudarshan1977,Itano1990,FacchiPascazio2008}. In superconducting circuits the quantum Zeno effect is well established, with the strength and timing of the measurements set by the qubit control gate sequence~\cite{Slichter2016,Harrington2017}. In these experiments, and in the theory behind them, the measurement times are classical parameters programmed into a controller. Here we ask what happens when two measurement schedules are instead placed in a coherent superposition. As far as we are aware, this type of quantum Zeno protocol has not previously been analyzed.

This question is motivated by relational approaches to time. In the Page--Wootters construction, time is the reading of a quantum clock, and the evolution of everything else is recovered by conditioning a stationary global state on that reading~\cite{PageWootters1983,Wootters1984,GLM2015,HohnSmithLock2021}. If the clock is itself quantum, nothing requires a measurement to occur at a single, definite clock reading. This becomes significant when clocks interact gravitationally; whether an event is localized in time can then depend on which clock is used as the reference~\cite{CastroRuiz2020}. More broadly, detectors can be switched on in a superposition of times~\cite{Henderson2020}, and the order of two operations can be coherently controlled, as in the quantum switch~\cite{Oreshkov2012,Chiribella2013,Procopio2015,Goswami2018}. These ideas are closely related, since superposing when events happen can superpose their order~\cite{Zych2019}. A related spatial effect shows that coherent control over trajectories through different environments can freeze open-system dynamics, producing a counterpart of the quantum Zeno effect in the spatial domain~\cite{Lin2022}. Because the quantum Zeno effect depends sensitively on measurement timing, it provides a natural way to ask what a superposed measurement schedule does.

Coherent control of quantum channels has previously been studied, and it has been shown that interference between alternative processes gives statistics distinct from those of a classical mixture~\cite{Abbott2020}. The work presented here applies coherent schedule superposition to the measurement timing of the quantum Zeno effect, investigating what happens when two temporally shifted measurement schedules act on a driven qubit. We derive an exact relation that gives a surprising result, where conditioned on one outcome of the schedule-qubit readout, the probability of finding the system in its initial state is the complement of the corresponding standard Zeno probability. This leads to an effect where, as increasingly frequent recordings freeze the reference evolution, the selected branch approaches the opposite state. Our use of a physical memory register provides an independent test: measuring the joint parity of the memories distinguishes a coherent schedule from a classical or dephased one, without relying on the final state of the system qubit.

Here we study this question for a single driven qubit. A system qubit $S$ is driven at Rabi frequency $\Omega$ and recorded $N$ times in the computational basis, each time by a CNOT onto a fresh memory qubit. A schedule qubit $Q$ decides whether an extra drive segment of duration $\delta$ is applied before or after the whole recording sequence, so the two branches contain the same sequence shifted by $\delta$ relative to the drive. This shift is equivalent to exchanging the order of the sequence and one segment of the drive. The circuit reproduces the branch structure of a quantum switch between that segment and the recording sequence. We implement these branches using an early and a late controlled rotation in a fixed-order gate sequence, without requiring indefinite causal order. The subsequent readout of $Q$ in the $\ket{\pm}_Q$ basis then projects onto the anticommutator and commutator of the two operations, the same property the switch uses to decide whether two gates commute or anticommute~\cite{Chiribella2012,Procopio2015}.

For the ideal circuit we find two results. First, the overlap of the two branches, which sets the fringe visibility of $Q$, is $\mathcal{V}=\cos^2(\Omega\delta/2)$ for any number $N\geq1$ of intermediate measurement records and any spacing between them. In other words, adding records does not make the schedules more distinguishable, because the branches differ only at the ends of the sequence. Without the records, $\mathcal{V}=1$. Second, the minus outcome of $Q$ only retains the $\sigma_x$ part of the offset rotation. It heralds that the offset flipped $S$, either before or after the recording sequence, and given that outcome the probability of finding $S$ in its initial state is exactly $1-P_0^{(N)}$, where $P_0^{(N)}$ is the corresponding probability in a standard Zeno run with zero offset. Adding more records drives the selected probability to zero while the reference goes to one. The state conditioned on the minus outcome does not depend on $\delta$, which only sets how often it occurs. By contrast, if $Q$ is ignored entirely, the statistics of $S$ reduce to those of a classical, equal-weight mixture of the two schedules. The statistics of $S$ alone therefore cannot distinguish a coherent schedule from a classical mixture; the distinction appears only when $Q$ is read out.

In our protocol the memories are physical qubits, and they do more than record. Given the minus outcome, the memory register is left in an eigenstate of its joint $X$-basis parity with an eigenvalue of $-1$, for any number and spacing of the records, while a classical or dephased schedule gives zero parity. Reading the memories tests the coherence between the two recorded histories directly. 

Without the memories, however, the statistics of $Q$ and $S$ depend only on the dephasing that each record applies to $S$. A variant of the circuit with mid-circuit measurements of $S$ in place of the memories reproduces those statistics and provides a cross-check, but it gives no access to the records themselves. 

The paper is organized as follows. Section~\ref{sec:setup} defines the circuit and its two measurement schedules; Sec.~\ref{sec:reference} reviews the standard Zeno reference probabilities;  Sec.~\ref{sec:superposed} derives the interference visibility and the conditional system probabilities that distinguish a coherent schedule from a classical or dephased one; Sec.~\ref{sec:experiment} proposes a circuit implementation that includes four experimental signatures, and Sec.~\ref{sec:discussion} discusses the relational interpretation and concludes.

These predictions follow from standard quantum mechanics, and they do not distinguish a relational description of time from an external one. The Page--Wootters picture contributes the reason to treat the schedule as a quantum degree of freedom, together with a description in which the whole protocol, readout included, is a single stationary history state, which we discuss in Appendix~\ref{app:embedding}. 

\section{Physical Setup and Measurement Schedules}
\label{sec:setup}

Our protocol uses a driven system qubit \(S\), a schedule qubit \(Q\), and a register of memory qubits \(M_1,\ldots,M_N\). Each memory interacts with the system once and stores a record of its computational-basis state. The schedule qubit controls whether an additional rotation of the system occurs before or after the recording sequence.

We prepare the system and all the memories in their ground states, with no initial correlations between them. We also assume that each unused memory stays in its ready state until its recording interaction. 

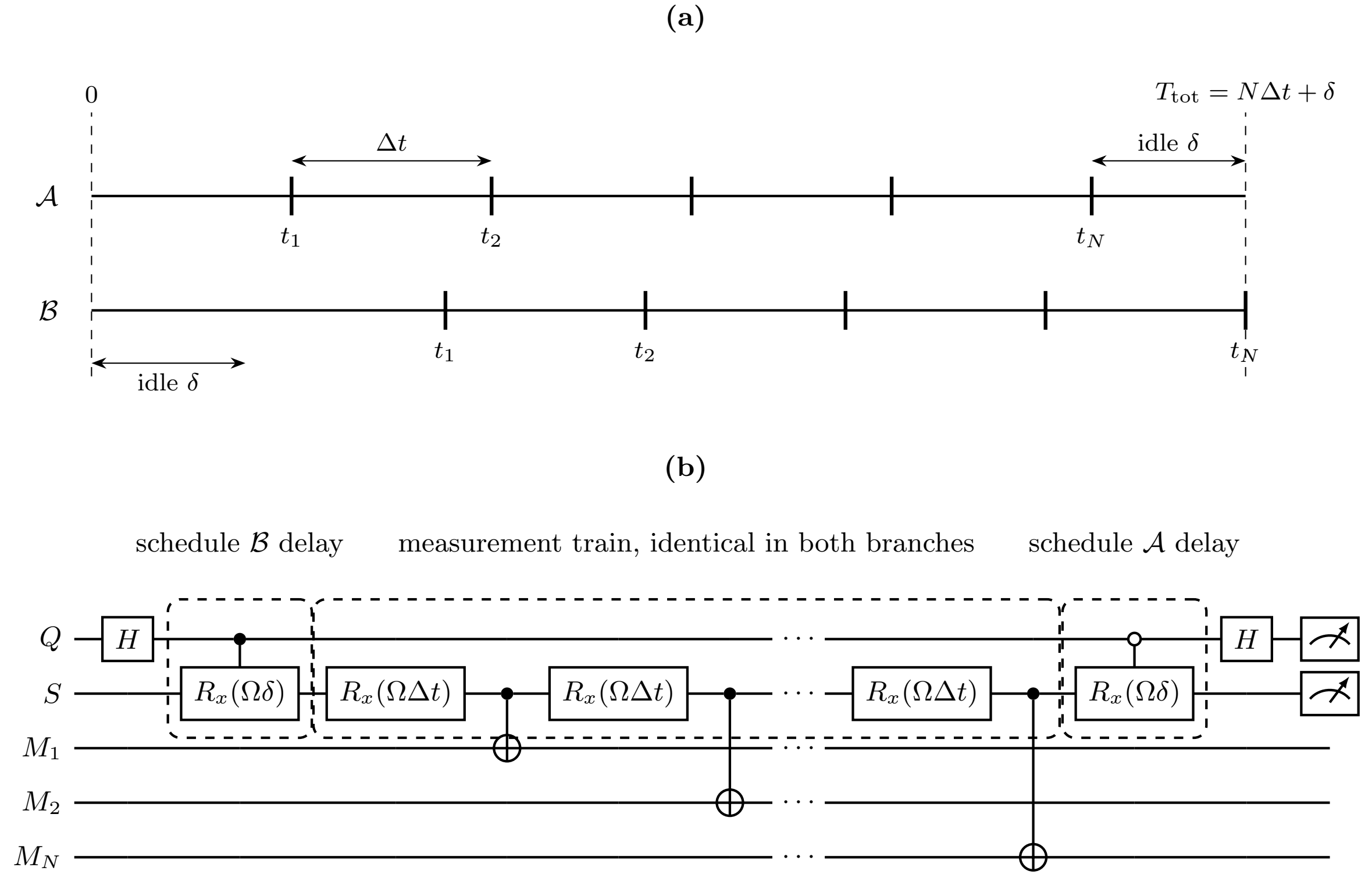
\begin{figure*}[t]
\noindent{\bfseries (a)}\par\vspace{2pt}
\begin{center}
\begin{tikzpicture}[x=1.75cm,y=1cm,>=Stealth]
\draw[thick] (0,1.3) -- (7.5,1.3); \node[left] at (-0.15,1.3) {$\mathcal{A}$};
\draw[thick] (0,0) -- (7.5,0);     \node[left] at (-0.15,0) {$\mathcal{B}$};
\foreach \k in {1,...,5} { \draw[very thick] (\k*1.3,1.08) -- (\k*1.3,1.52); }
\foreach \k in {1,...,5} { \draw[very thick] (\k*1.3+1,-0.22) -- (\k*1.3+1,0.22); }
\node[below,font=\footnotesize] at (1.3,1.05) {$t_1$};
\node[below,font=\footnotesize] at (2.6,1.05) {$t_2$};
\node[below,font=\footnotesize] at (6.5,1.05) {$t_N$};
\node[below,font=\footnotesize] at (2.3,-0.25) {$t_1$};
\node[below,font=\footnotesize] at (3.6,-0.25) {$t_2$};
\node[below,font=\footnotesize] at (7.5,-0.25) {$t_N$};
\draw[dashed] (0,-0.75) -- (0,2.25);   \node[above,font=\footnotesize] at (0,2.25) {$0$};
\draw[dashed] (7.5,-0.75) -- (7.5,2.25); \node[above,font=\footnotesize] at (7.5,2.25) {$T_{\rm tot}=N\Delta t+\delta$};
\draw[<->] (1.3,1.7) -- (2.6,1.7) node[midway,above,font=\footnotesize] {$\Delta t$};
\draw[<->] (6.5,1.7) -- (7.5,1.7) node[midway,above,font=\footnotesize] {idle $\delta$};
\draw[<->] (0,-0.6) -- (1.0,-0.6) node[midway,below,font=\footnotesize] {idle $\delta$};
\end{tikzpicture}
\end{center}
\vspace{4pt}
\noindent{\bfseries (b)}\par\vspace{2pt}
\begin{center}
\begin{quantikz}[row sep={0.62cm,between origins}, column sep=0.32cm]
\lstick{$Q$}   & \gate{H} & \ctrl{1}\gategroup[2,steps=1,style={dashed,rounded corners,inner xsep=1pt,inner ysep=2.5pt},background,label style={label position=above,anchor=south,yshift=0.15cm}]{schedule $\mathcal{B}$ delay} & \qw\gategroup[2,steps=7,style={dashed,rounded corners,inner xsep=1pt,inner ysep=2.5pt},background,label style={label position=above,anchor=south,yshift=0.15cm}]{measurement train, identical in both branches} & \qw & \qw & \qw & \ \cdots\ \qw & \qw & \qw & \octrl{1}\gategroup[2,steps=1,style={dashed,rounded corners,inner xsep=1pt,inner ysep=2.5pt},background,label style={label position=above,anchor=south,yshift=0.15cm}]{schedule $\mathcal{A}$ delay} & \gate{H} & \meter{} \\
\lstick{$S$}   & \qw & \gate{R_x(\Omega\delta)} & \gate{R_x(\Omega\Delta t)} & \ctrl{1} & \gate{R_x(\Omega\Delta t)} & \ctrl{2} & \ \cdots\ \qw & \gate{R_x(\Omega\Delta t)} & \ctrl{3} & \gate{R_x(\Omega\delta)} & \qw & \meter{} \\
\lstick{$M_1$} & \qw & \qw & \qw & \targ{} & \qw & \qw & \ \cdots\ \qw & \qw & \qw & \qw & \qw & \qw \\
\lstick{$M_2$} & \qw & \qw & \qw & \qw & \qw & \targ{} & \ \cdots\ \qw & \qw & \qw & \qw & \qw & \qw \\
\lstick{$M_N$} & \qw & \qw & \qw & \qw & \qw & \qw & \ \cdots\ \qw & \qw & \targ{} & \qw & \qw & \qw
\end{quantikz}
\end{center}
\caption{(a) The two measurement schedules of Eq.~\eqref{eq:schedules}. Both contain $N$ measurements with spacing $\Delta t$ and both end at $T_{\rm tot}=N\Delta t+\delta$. Schedule $\mathcal{A}$ idles for $\delta$ after its last measurement and schedule $\mathcal{B}$ idles for $\delta$ before its first. (b) The circuit that realizes their superposition. The first Hadamard prepares the schedule qubit $Q$ in $\ket{+}_Q$. The delay rotation $R_x(\Omega\delta)$ on $S$ is applied before the measurement train when $Q$ is in $\ket{B}_Q$ (filled control) and after it when $Q$ is in $\ket{A}_Q$ (open control). The train of drive rotations and CNOT gates into the memory register is identical in both branches and is not controlled by $Q$. A final Hadamard on $Q$ followed by readout of $Q$ and $S$ implements the readout described later in Eq.~\eqref{eq:readoutState}. The memory register remains unread when measuring the conditional system probabilities; the additional record-parity test described later in Section~\ref{sec:experiment} measures it in the $X$ basis.}
\label{fig:circuit}
\end{figure*}

Each recording interaction is a CNOT with the system as the control and a fresh memory as the target. This can entangle the system with the memory while keeping the combined ideal state pure. The memories are not measured during the sequence; they remain entangled with $S$ until the final readout. To obtain the conditional system probabilities, we measure the schedule qubit and the system, leaving the memories unread (Fig.~\ref{fig:circuit}). For the additional record-parity test described in Sec.~\ref{sec:experiment}, we also measure the memories in the $X$ basis.

The schedule qubit does not act as the clock that orders the gates. Instead, laboratory controls apply the chosen rotations and CNOTs in a sequence that is fixed in advance. The stationary history construction in Appendix B represents this same sequence using an additional clock register, but the proposed experiment does not require preparing that register.

\subsection{Driven System}
We take $S$ to be a driven qubit, initially prepared in $\ket{0}_S$. Using units where $\hbar=1$, its drive Hamiltonian is
\begin{equation}
\hat H_S=\frac{\Omega}{2}\sigma_x,
\label{eq:HS}
\end{equation}
where $\Omega$ is the applied drive strength, setting the rotation rate between the computational-basis states.

Before adding the recording interactions, evolution for an interval $t$ gives
\begin{equation}
\begin{aligned}
\ket{\psi_S(t)}
&=e^{-i\hat H_St}\ket{0}_S\\
&=\cos\left(\frac{\Omega t}{2}\right)\ket{0}_S
-i\sin\left(\frac{\Omega t}{2}\right)\ket{1}_S.
\end{aligned}
\label{eq:baselineState}
\end{equation}
The probability of finding the system in its initial state is therefore
\begin{equation}
P_0(t)
=\left|\braket{0|\psi_S(t)}\right|^2
=\cos^2\left(\frac{\Omega t}{2}\right).
\label{eq:baselineProbability}
\end{equation}

In the circuit implementation, we reproduce this evolution with a rotation $R_x(\Omega t)$. The model interval $t$ and drive strength $\Omega$ used together set the rotation angle $\theta=\Omega t$. A control pulse can implement the same rotation gate using a different drive strength and duration. Because the laboratory controls set the entire gate sequence in advance, there is no need to measure the system's intermediate state to decide when to apply the next gate.

\subsection{Recording Interactions}

We represent each measurement by an interaction between $S$ and a fresh memory qubit, following the memory construction of Pinto Couto \textit{et al.}~\cite{PintoCouto2026}. Suppose the system is in the state $\alpha\ket{0}_S+\beta\ket{1}_S$ and the first memory is ready in $\ket{0}_{M_1}$. A CNOT from $S$ to $M_1$ gives
\begin{equation}
\begin{aligned}
&\left(\alpha\ket{0}_S+\beta\ket{1}_S\right)\ket{0}_{M_1}\\
&\qquad\longrightarrow
\alpha\ket{0}_S\ket{0}_{M_1}
+\beta\ket{1}_S\ket{1}_{M_1}.
\end{aligned}
\label{eq:measurementInteraction}
\end{equation}
The memory now holds a record of the system's computational-basis state at this interaction. If we look only at $S$, its reduced state has no coherence between $\ket{0}_S$ and $\ket{1}_S$. The combined system and memory still retain the quantum information, and the interaction is unitary.

For the $k$th recording interaction, we write
\begin{equation}
\hat U_{\mathrm{meas}}^{(k)}
=\mathrm{CNOT}_{S\rightarrow M_k}.
\end{equation}
Between recordings, the system undergoes the drive rotation
\begin{equation}
\hat U_S(\Delta t)
=e^{-i\hat H_S\Delta t}
=R_x(\Omega\Delta t).
\end{equation}
Each rotation is followed by a CNOT into the next unused memory. After $N$ such steps, the state is
\begin{equation}
\begin{aligned}
\ket{\psi_{SM}^{(N)}}
&=\hat U_{\mathrm{meas}}^{(N)}
\hat U_S(\Delta t)\cdots\\
&\qquad\times
\hat U_{\mathrm{meas}}^{(1)}
\hat U_S(\Delta t)
\ket{0}_S\ket{0}_M,
\end{aligned}
\label{eq:fullZenoSequence}
\end{equation}
where $\ket{0}_M$ means that all $N$ memories start in their ground states. We write $T=N\Delta t$ for the total drive interval within this recording sequence. We will introduce the additional schedule offset $\delta$ in the next section.

Each memory interacts with $S$ only once. We do not reset or measure the memories between steps. Tracing over the memories then gives the same statistics as projective measurements on $S$.

\subsection{Two Measurement Schedules}

We now consider two measurement schedules, $\mathcal{A}$ and $\mathcal{B}$, that differ only by a fixed offset $\delta$ of the entire measurement train,
\begin{equation}
\mathcal{A}: \;t_k = k\,\Delta t, \qquad \mathcal{B}: \;t_k = k\,\Delta t + \delta, \qquad k=1,\dots,N 
\label{eq:schedules}
\end{equation}
as shown in Fig.~\ref{fig:circuit}(a). The offset $\delta$ is independent of the measurement spacing $\Delta t$. We do not assume $\delta<\Delta t$, and the results below hold for arbitrary $\delta$.
Both schedules are read out at the same final time
\begin{equation}
T_{\mathrm{tot}} = N\Delta t + \delta ,
\end{equation}
so schedule $\mathcal{A}$ leaves an extra idle interval $\delta$ after its last measurement, and schedule $\mathcal{B}$ leaves an idle interval $\delta$ before its first one. Everything else about the two circuits is identical, including the memory register they are written into.

To superpose these two schedules, we introduce a two-dimensional schedule qubit $Q$ with orthogonal states $\ket{A}_Q$ and $\ket{B}_Q$, where $\ket{A}_Q$ selects schedule $\mathcal{A}$ and $\ket{B}_Q$ selects schedule $\mathcal{B}$. We prepare it in the superposition
\begin{equation}
\ket{\chi}_Q=\tfrac{1}{\sqrt2}\left(\ket{A}_Q+\ket{B}_Q\right).
\end{equation}
We identify $\ket{A}_Q$ with the computational-basis state $\ket{0}_Q$ and $\ket{B}_Q$ with $\ket{1}_Q$. The circuit first applies $R_x(\Omega\delta)$ to $S$ conditioned on $Q$ being in $\ket{B}_Q$. It then applies the recording sequence, followed by $R_x(\Omega\delta)$ conditioned on $Q$ being in $\ket{A}_Q$. The drive rotations and recording CNOTs within the sequence are identical in both branches and are not controlled by $Q$.

For the ideal drive and recording gates considered here, these two placements of the offset rotation reproduce the final branch states of the shifted schedules in Fig.~\ref{fig:circuit}(a). In the circuit implementation, the laboratory executes the gate sequence shown in Fig.~\ref{fig:circuit}(b). 

The schedule qubit coherently controls the placement of the offset rotation within that sequence. It selects between two schedules rather than labeling successive stages of the experiment. This is distinct from both the continuous reference clock $C$ used in the Page--Wootters background and the finite history clock $\mathcal{T}$ introduced in Appendix~\ref{app:embedding}.

Starting with $Q$ in $\ket{\chi}_Q$ gives the joint state
\begin{equation}
\ket{\Psi}=\frac{1}{\sqrt2}\Bigl(\ket{A}_Q\otimes\ket{\psi_{SM}^{\mathcal{A}}}+\ket{B}_Q\otimes\ket{\psi_{SM}^{\mathcal{B}}}\Bigr),
\label{eq:superposedHistory}
\end{equation}

where $\ket{\psi_{SM}^{\mathcal{A}}}$ and $\ket{\psi_{SM}^{\mathcal{B}}}$ are the final system--memory states for the two schedules. Schedule $\mathcal{A}$ applies the offset rotation after the recording sequence in Eq.~\eqref{eq:fullZenoSequence}, while schedule $\mathcal{B}$ applies it before. Eq.~\eqref{eq:superposedHistory} is the joint circuit state before the schedule readout. Appendix~\ref{app:embedding} shows how the full sequence can be represented as a stationary history.

\subsection{Final Readout}

After the gate sequence, we apply a Hadamard to $Q$ and measure it in the computational basis. An outcome of zero corresponds to the symmetric schedule superposition, while an outcome of one corresponds to the antisymmetric superposition. We refer to these as the plus and minus outcomes. In addition to $Q$, we also measure $S$ in the computational basis. For these system-statistics measurements, the memory register is left unread.

We group the experimental runs by the outcome recorded for $Q$. Within each group, we count the fraction of runs where $S$ is found in zero. These fractions give the conditional probabilities $P_0^{(N,+)}$ and $P_0^{(N,-)}$. We also count how often each schedule outcome occurs, giving $p_+$ and $p_-$. If we ignore the schedule outcome and record only whether $S$ is found in its initial state, we obtain the unread probability.

To include the schedule readout in the stationary history description, we represent its detector by a readout memory $W$. This memory stores the result of the schedule measurement. In the proposed implementation, the laboratory detector supplies this record, so $W$ does not require an additional processor qubit. The explicit readout model and its probabilities are given in Sec.~\ref{sec:differentoutcomes}.

\section{Standard Zeno Reference}
\label{sec:reference}
Repeated recording interactions suppress transitions away from the initial state when the intervals between them become short, producing the quantum Zeno effect~\cite{MisraSudarshan1977,Itano1990}. The standard quantum Zeno survival probability asks whether the system remained in zero at every step. Our main observable is different: whether the system is in zero at the end, regardless of what the memories recorded along the way.

In the standard quantum Zeno effect protocol, we consider the record where every memory contains zero. The probability of staying in $\ket{0}_S$ during one drive interval is $\cos^2(\Omega\Delta t/2)$. The probability of obtaining zero at all $N$ recordings is therefore
\begin{equation}
P_{\mathrm{surv}}^{(N)}
=\left[\cos^2\left(\frac{\Omega\Delta t}{2}\right)\right]^N.
\label{eq:zenoSurvival}
\end{equation}
This is the probability of the memory record $\ket{00\cdots0}_M$. Measuring it requires reading the memories and counting the runs where every recorded bit is zero.

Our main observable is instead the probability of finding $S$ in $\ket{0}_S$ at the end, with the memories left unread. This includes records where the system was recorded in one at an earlier step and returned to zero later. It is not the same as the probability of an all-zero record.

To calculate this final-zero probability, we follow the difference between the populations of $\ket{0}_S$ and $\ket{1}_S$. After each recording interaction, the reduced state of $S$ is diagonal in the computational basis, with population in both $\ket{0}_S$ and $\ket{1}_S$ in general. Each subsequent drive rotation followed by a recording then multiplies the population difference by $\cos(\Omega\Delta t)$. Starting from $\ket{0}_S$, the initial population difference is one. After $N$ steps, this gives
\begin{equation}
\begin{aligned}
P_0^{(N)}(T)
&=\frac{1}{2}\left[1+\cos^N\left(\Omega\Delta t\right)\right],\\
\Delta t&=\frac{T}{N}.
\end{aligned}
\label{eq:zenoUnconditional}
\end{equation}

Both probabilities approach one as $N$ increases with $T$ fixed, but they count different outcomes. Throughout the results below, $P_0^{(N)}(T)$ denotes the final-zero probability in a separate reference run with zero schedule offset and the recording interactions enabled. This is the reference used in the complement relation for the superposed schedules. It only requires measuring $S$ at the end.

\section{Interference and Conditional Zeno Dynamics}
\label{sec:superposed}

We now calculate the interference between the two schedules and the system's final-zero probability for each schedule-readout outcome. We compare these probabilities with the standard Zeno reference and with the result obtained when the schedule outcome is ignored.

\subsection{Interference Between Measurement Schedules}

We first calculate how much coherence remains between the two measurement schedules after the sequence, by comparing the final $S+M$ states produced by schedules $\mathcal{A}$ and $\mathcal{B}$. Their overlap measures how similar these two quantum states are,
\begin{equation}
\mathcal{V}=\left|\braket{\psi_{SM}^{\mathcal{B}}|\psi_{SM}^{\mathcal{A}}}\right| .
\label{eq:visdef}
\end{equation}
An overlap of $1$ means that the $S+M$ states associated with the two schedule branches are identical, as they are when $\delta=0$, while an overlap of $0$ means that the two branch states are orthogonal and therefore perfectly distinguishable. This is the usual which-path tradeoff: information about which schedule occurred is carried by the joint $S$+$M$ state, so the more distinguishable the two branch states become, the less coherence remains in $Q$.

Experimentally, this remaining coherence appears as the fringe contrast of a Ramsey measurement on $Q$, since the coherence between the two branches is proportional to the overlap $\braket{\psi_{SM}^{\mathcal{B}}|\psi_{SM}^{\mathcal{A}}}$. 

To calculate this overlap, we keep track of the different measurement records stored in the memory register. We label one possible record by $\vec m=(m_1,\dots,m_N)$, where each $m_k$ is either $0$ or $1$ and records the state of $S$ at the $k$th measurement interaction. Because $\hat U_{\mathrm{meas}}^{(k)}$ records whether $S$ is in $\ket{0}_S$ or $\ket{1}_S$ in the memory qubit $M_k$, the last memory value $m_N$ is correlated with the state $\ket{m_N}_S$ immediately after the final measurement.

For schedule $\mathcal{A}$, the final state can be written as
\begin{equation}
\ket{\psi_{SM}^{\mathcal{A}}}=\sum_{\vec m}c^{\mathcal{A}}_{\vec m}\; \bigl(\hat U_S(\delta)\ket{m_N}_S\bigr)\otimes\ket{\vec m}_M,
\label{eq:finalstate}
\end{equation}
where
\begin{equation}
c^{\mathcal{A}}_{\vec m}=\braket{m_N|\hat U_S(\Delta t)|m_{N-1}}
\cdots\braket{m_1|\hat U_S(\Delta t)|0}
\end{equation}
is the amplitude for the record $\vec m$ in schedule $\mathcal{A}$. For schedule $\mathcal{B}$,
\begin{equation}
\begin{aligned}
c^{\mathcal{B}}_{\vec m}
&=\braket{m_N|\hat U_S(\Delta t)|m_{N-1}}\cdots\\
&\quad\times
\braket{m_2|\hat U_S(\Delta t)|m_1}
\braket{m_1|\hat U_S(\Delta t+\delta)|0}.
\end{aligned}
\end{equation}
So the two amplitudes differ only in the first evolution interval, while schedule $\mathcal{A}$ instead has the additional free-evolution interval $\delta$ after its final measurement.

The overlap between the two final states is therefore
\begin{equation}
\braket{\psi_{SM}^{\mathcal{B}}|\psi_{SM}^{\mathcal{A}}}=\sum_{\vec m}\left(c^{\mathcal{B}}_{\vec m}\right)^* c^{\mathcal{A}}_{\vec m}\;\braket{m_N|\hat U_S(\delta)|m_N}.
\label{eq:finaloverlap}
\end{equation}

For either possible final value $m_N=0$ or $m_N=1$,
\begin{equation}
\braket{m_N|\hat U_S(\delta)|m_N}=\cos\left(\frac{\Omega\delta}{2}\right).
\end{equation}
This is the same for both values of $m_N$ because the $\sigma_x$ Hamiltonian exchanges $\ket{0}_S$ and $\ket{1}_S$ symmetrically, giving the same amplitude to remain in either state after the interval $\delta$. This factor is the same for every memory record and can be taken outside the sum.

The remaining sum of Eq.~\eqref{eq:finaloverlap} simplifies because, apart from the offset at the beginning and end of the sequence, the two schedules have exactly the same intervals $\Delta t$. When we sum over all possible values recorded by the intermediate memory qubits, the probabilities for each possible transition add to one,
\begin{equation}
\sum_{m_k}\left|\braket{m_k|\hat U_S(\Delta t)|m_{k-1}}\right|^2=1.
\end{equation}
As a result, the contributions from the identical intermediate intervals each sum to unity and drop out of the overlap one by one. The only remaining difference comes from the first interval,
\begin{align}
&\sum_{m_1}\left(\braket{m_1|\hat U_S(\Delta t+\delta)|0}\right)^*\braket{m_1|\hat U_S(\Delta t)|0}\nonumber\\
&=\bra{0}\hat U_S^\dagger(\Delta t+\delta)\left(\sum_{m_1}\ket{m_1}\!\bra{m_1}\right)\hat U_S(\Delta t)\ket{0}\nonumber\\
&=\bra{0}\hat U_S(-\Delta t-\delta)\hat U_S(\Delta t)\ket{0}\nonumber\\
&=\braket{0|\hat U_S(-\delta)|0}.
\end{align}
This gives another factor
\begin{equation}\braket{0|\hat U_S(-\delta)|0}=\cos\left(\frac{\Omega\delta}{2}\right).
\end{equation}

Combining the two factors gives the schedule visibility
\begin{equation}
\boxed{\mathcal{V}=\cos^{2}\left(\frac{\Omega\delta}{2}\right)}
\label{eq:visibility}
\end{equation}
so the coherence between the two measurement schedules depends only on the offset $\delta$. It does not depend on the number of measurements $N$, the spacing $\Delta t$ between them, or the total duration of the sequence. 

Eq.~\eqref{eq:visibility} applies only when the measurement interactions are present. Without the measurement interactions, the only difference between the two schedules is that the extra free-evolution interval $\delta$ occurs before the measurement sequence in $\mathcal{B}$ and after it in $\mathcal{A}$. Since $S$ undergoes the same total amount of free evolution in either case, the two schedules produce the same final state of $S$, so their overlap is
\begin{equation}
\left|\braket{\psi_S^{\mathcal{B}}|\psi_S^{\mathcal{A}}}\right|=1.
\end{equation}
Therefore, $\mathcal{V}=1$ for any value of $\delta$ when the measurement interactions are absent. Coherence begins to drop only when those interactions make the two $S+M$ branch states distinguishable enough to carry information about which schedule occurred.

Also, increasing the number of intermediate measurements $N$ does not further reduce $\mathcal{V}$. The intermediate parts of the two schedules are identical. They differ only where the extra interval $\delta$ appears, before the measurement train in schedule $\mathcal{B}$ or after it in schedule $\mathcal{A}$. The intermediate memory records carry no additional information about which schedule occurred, so the entire distinction between the two branches is held in the first memory record together with the final state of $S$. 

This cancellation of the intermediate records is not itself a consequence of the Zeno limit. It follows because the intermediate evolution and measurement interactions are identical in the two branches, so summing over each intermediate memory record gives unity by completeness.

\subsection{Zeno Dynamics for Different Readout Outcomes}
\label{sec:differentoutcomes}

The second and more striking observable is what happens to $S$ when $Q$ is read in the superposition basis
\begin{equation}
\ket{\pm}_Q=\tfrac{1}{\sqrt2}\left(\ket{A}_Q\pm\ket{B}_Q\right).
\end{equation}
We treat this readout the same way we treated the measurements of $S$, as unitary correlation with a memory rather than as a projection applied from outside. A Hadamard gate on $Q$ followed by a CNOT gate from $Q$ to an ancilla $W$ prepared in $\ket{0}_W$ maps $\ket{+}_Q\to\ket{A}_Q\ket{0}_W$ and $\ket{-}_Q\to\ket{B}_Q\ket{1}_W$, so Eq.~\eqref{eq:superposedHistory} becomes
\begin{align}
\ket{\Psi'}=\tfrac12\Bigl[
&\ket{A}_Q\ket{0}_W
\bigl(\ket{\psi_{SM}^{\mathcal{A}}}+\ket{\psi_{SM}^{\mathcal{B}}}\bigr)\nonumber\\
+&\ket{B}_Q\ket{1}_W\bigl(\ket{\psi_{SM}^{\mathcal{A}}}-\ket{\psi_{SM}^{\mathcal{B}}}\bigr)\Bigr].
\label{eq:readoutState}
\end{align}
The record $W=0$ is correlated with the $\ket{+}_Q$ outcome and $W=1$ with the $\ket{-}_Q$ outcome, and the unnormalized $S+M$ states carrying the two records are $\tfrac12(\ket{\psi_{SM}^{\mathcal{A}}}\pm\ket{\psi_{SM}^{\mathcal{B}}})$. Their squared norms give the probabilities of the two records,
\begin{equation}
p_\pm=\tfrac12\left(1\pm\mathcal{V}\right)=\tfrac12\left[1\pm\cos^{2}\!\left(\frac{\Omega\delta}{2}\right)\right].
\label{eq:ppm}
\end{equation}
Writing $\hat P_0=\ket{0}\!\bra{0}_S\otimes\Id_M$, the unnormalized $S+M$ states associated with the two readout records are
\begin{equation}
\ket{\phi_\pm}=\frac{1}{2}\left(\ket{\psi_{SM}^{\mathcal{A}}}\pm\ket{\psi_{SM}^{\mathcal{B}}}\right).
\end{equation}
The joint probability of obtaining the corresponding record and finding $S$ in its initial state is therefore
\begin{align}
\bra{\phi_\pm}\hat P_0\ket{\phi_\pm}=\frac{1}{4}\Bigl[P_0^{\mathcal{A}}+P_0^{\mathcal{B}}\pm2\,\mathrm{Re}\braket{\psi_{SM}^{\mathcal{B}}|\hat P_0|\psi_{SM}^{\mathcal{A}}}\Bigr].
\end{align}
Dividing this joint probability by the probability $p_\pm=\tfrac12(1\pm\mathcal V)$ of obtaining the corresponding readout record gives the conditional probability of finding $S$ in its initial state,
\begin{equation}
P_0^{(N,\pm)}=\frac{P_0^{\mathcal{A}}+P_0^{\mathcal{B}}\pm 2\,\mathrm{Re}\braket{\psi_{SM}^{\mathcal{B}}|\hat P_0|\psi_{SM}^{\mathcal{A}}}}{2\left(1\pm\mathcal{V}\right)} .
\label{eq:clockOutcomeGeneral}
\end{equation}

The individual branch probabilities can be obtained from the same Bloch-vector contraction used for Eq.~\eqref{eq:zenoUnconditional}. Each measurement interaction removes the transverse coherence, so an interval $\Delta t$ followed by a measurement multiplies the population difference by $\cos(\Omega\Delta t)$. For schedule $\mathcal{A}$, this occurs $N$ times, after which the final free interval $\delta$ multiplies the remaining population difference by $\cos(\Omega\delta)$. This gives
\begin{equation}
P_0^{\mathcal{A}}=\tfrac12\left[1+\cos^{N}(\Omega\Delta t)\cos(\Omega\delta)\right].
\end{equation}

For schedule $\mathcal{B}$, the first free-evolution interval has duration $\Delta t+\delta$, giving a factor $\cos[\Omega(\Delta t+\delta)]$, while the remaining $N-1$ measured intervals each contribute a factor $\cos(\Omega\Delta t)$. Therefore,
\begin{equation}
P_0^{\mathcal{B}}=\tfrac12\left[1+\cos^{N-1}(\Omega\Delta t)\cos\bigl(\Omega(\Delta t+\delta)\bigr)\right].
\end{equation}

Ignoring the readout outcome is equivalent to tracing over the schedule-qubit readout. The interference terms between the two schedule branches then cancel, leaving the equal incoherent mixture of schedules $\mathcal{A}$ and $\mathcal{B}$. This gives
\begin{equation}
\overline{P_0}^{(N)}=p_+P_0^{(N,+)}+p_-P_0^{(N,-)}=\tfrac12\left(P_0^{\mathcal{A}}+P_0^{\mathcal{B}}\right),
\label{eq:marginal}
\end{equation}
which is not the same as Eq.~\eqref{eq:zenoUnconditional} once $\delta\neq0$. At fixed $T$, $\Delta t=T/N$, and therefore
\begin{equation}
\begin{aligned}
\cos^N(\Omega\Delta t)
&=\cos^N\left(\frac{\Omega T}{N}\right)\\
&\simeq\left(1-\frac{\Omega^2T^2}{2N^2}\right)^N\\
&\longrightarrow 1
\qquad (N\rightarrow\infty).
\end{aligned}
\end{equation}
At the same time, $\Omega\Delta t\rightarrow0$, so
\begin{equation}
\lim_{N\rightarrow\infty}\overline{P_0}^{(N)}=\tfrac12\left[1+\cos(\Omega\delta)\right]=\cos^{2}\!\left(\frac{\Omega\delta}{2}\right)=\mathcal{V}.
\label{eq:margLimit}
\end{equation}
The offset interval $\delta$ is not placed between two recording interactions in either branch. It undergoes free evolution regardless of $N$, and the resulting overlap is exactly the schedule visibility. The unread probability approaches one whenever
$\Omega\delta=2\pi k$ for an integer $k$, including $\delta=0$. At these offsets, $\mathcal{V}=1$ and the minus outcome does not occur.

The cross term in Eq.~\eqref{eq:clockOutcomeGeneral} can be evaluated using two properties of the recording sequence. Let $\hat X$ denote the sequence of Eq.~\eqref{eq:fullZenoSequence}, viewed as a map from $S$ to $S+M$, so that $\ket{\psi_{SM}^{\mathcal{A}}}=\hat U_S(\delta)\hat X\ket{0}_S$ and $\ket{\psi_{SM}^{\mathcal{B}}}=\hat X\hat U_S(\delta)\ket{0}_S$, and split the offset rotation as
\begin{equation}
\hat U_S(\delta)=\cos\left(\frac{\Omega\delta}{2}\right)\Id-i\sin\left(\frac{\Omega\delta}{2}\right)\sigma_x .
\label{eq:offsetSplit}
\end{equation}
First, $\sigma_x$ commutes with every drive rotation, and a CNOT turns a flip of its control into a flip of both qubits, so
\begin{equation}
\hat X\sigma_x=\sigma_x^{S}\hat\Sigma_M\hat X,\qquad \hat\Sigma_M=\bigotimes_{k=1}^{N}\sigma_x^{M_k}.
\label{eq:flipCovariance}
\end{equation}
Second, the last CNOT leaves $S$ equal to the record in $M_N$, so any state in which $S$ disagrees with $M_N$ is orthogonal to every output of $\hat X$. Together with $\hat X^\dagger\hat X=\Id$, this gives $\hat X^\dagger\hat U_S(\delta)\hat X=\cos(\Omega\delta/2)\,\Id$, and Eq.~\eqref{eq:visibility} follows in one line. The identity parts of Eq.~\eqref{eq:offsetSplit} are the same in both branches, so they cancel in the antisymmetric combination, leaving
\begin{equation}
\begin{aligned}
\ket{\phi_-}&=\tfrac12\bigl[\hat U_S(\delta),\hat X\bigr]\ket{0}_S\\
&=-\frac{i}{2}\sin\left(\frac{\Omega\delta}{2}\right)\Bigl(\sigma_x^{S}\hat X\ket{0}_S-\hat X\ket{1}_S\Bigr).
\end{aligned}
\label{eq:minusState}
\end{equation}
Its squared norm reproduces $p_-=\tfrac12\sin^2(\Omega\delta/2)$. The first term is the zero-offset sequence followed by a flip of $S$, and the second is a flip followed by the sequence. 

\begin{figure*}[t]
\subfloat[]{\includegraphics[width=246pt]{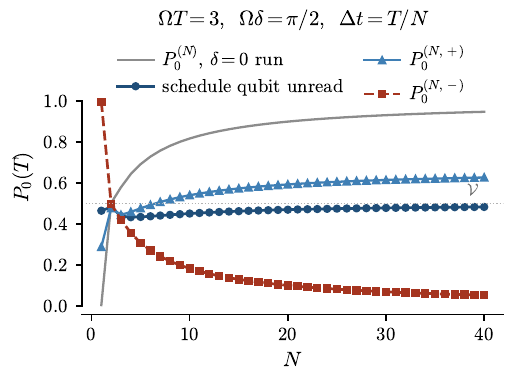}}
\hfill
\subfloat[]{\includegraphics[width=246pt]{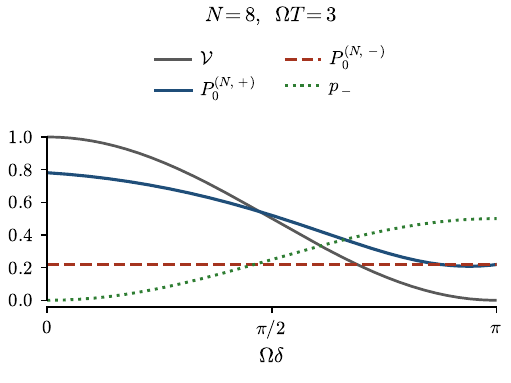}}
\caption{(a) Final-zero probability of $S$ at fixed total drive interval $\Omega T=3$ and offset $\Omega\delta=\pi/2$, as a function of the number of measurement interactions $N$, with $\Delta t=T/N$. The three marked curves come from a single dataset sorted by the readout outcome: the $\ket{-}_Q$ branch, the $\ket{+}_Q$ branch, and the unread mixture of Eq.~\eqref{eq:marginal}, which saturates at $\mathcal{V}=\tfrac12$ (dotted line) rather than at unity. The grey curve is the final-zero probability for the standard Zeno reference protocol of Eq.~\eqref{eq:zenoUnconditional}, measured in a separate $\delta=0$ reference run. The $\ket{-}_Q$ branch is its exact complement, Eq.~\eqref{eq:antizeno}. (b) Dependence on the schedule offset $\delta$ at $N=8$. The schedule visibility $\mathcal{V}$ is independent of $N$, the inverted branch $P_0^{(N,-)}$ is independent of $\delta$, and $P_0^{(N,+)}$ interpolates between the Zeno curve at $\delta\rightarrow0$ and the inverted value at $\Omega\delta=\pi$. The curve $p_-$ gives the fraction of shots where the inverted branch is obtained. The minus-selected curve is defined only where $p_->0$, and any value shown at $p_-=0$ represents its limiting value. We use $\Omega T=3$ to illustrate the separation between the zero-offset reference and the minus-selected result over the range of $N$ shown.}
\label{fig:main}
\end{figure*}

By Eq.~\eqref{eq:flipCovariance}, $\hat X\ket{1}_S=\sigma_x^S\hat\Sigma_M\hat X\ket{0}_S$, so both terms give $S$ the computational-basis statistics of the reference run with $\ket{0}_S$ and $\ket{1}_S$ exchanged. The two terms do not interfere in those statistics: the first has $S$ flipped relative to $M_N$, while in the second both are flipped together, so the two terms are orthogonal. This gives the central result of this paper,
\begin{equation}
\boxed{\;P_0^{(N,-)}(T)=\frac{1}{2}\left[1-\cos^{N}\!\left(\Omega\Delta t\right)\right]=1-P_0^{(N)}(T)\;}
\label{eq:antizeno}
\end{equation}
which holds whenever $p_->0$. This exact complement relation means that, as increasingly frequent recordings at fixed $T$ freeze the standard Zeno evolution, the minus-selected branch approaches the opposite state (Fig.~\ref{fig:main}). When $\Omega\delta=2\pi k$ for an integer $k$, the minus outcome never occurs, so its conditional probability is not defined. For the plus outcome, we obtain
\begin{equation}
\begin{aligned}
P_0^{(N,+)}
&=\frac{1}{2}+\frac{\cos^{N}(\Omega\Delta t)\cos^2\!\left(\frac{\Omega\delta}{2}\right)}{2\left[1+\cos^2\!\left(\frac{\Omega\delta}{2}\right)\right]}
\\
&\qquad+\frac{\cos^{N-1}(\Omega\Delta t)\cos\bigl(\Omega(\Delta t+\delta)\bigr)}{2\left[1+\cos^2\!\left(\frac{\Omega\delta}{2}\right)\right]}.
\end{aligned}
\label{eq:plusbranch}
\end{equation}

The visibility~\eqref{eq:visibility} depends on $\delta$, but the
heralded state~\eqref{eq:minusState} itself does not; because $\delta$
enters only through the probability $p_-$ of obtaining the minus
outcome, the state conditioned on that outcome has the same form for
every $\delta$.

These properties do not require uniform spacing. Neither property of $\hat X$ uses $\Delta t = T/N$, so Eq.~\eqref{eq:visibility} and
the relation $P_0^{(N,-)} = 1 - P_0^{(N)}$ hold for arbitrary measurement intervals,
with $P_0^{(N)}$ evaluated for the same intervals.

In the Zeno limit, where
$\hat X\ket{0}_S \rightarrow \ket{0}_S\ket{0\cdots0}_M$, the heralded
state approaches a GHZ-type superposition of the two uniform memory
records,
$\ket{1}_S\bigl(\ket{0\cdots0}_M - \ket{1\cdots1}_M\bigr)/\sqrt{2}$,
and its fidelity with that limiting state is exactly the all-zero
record probability $P_{\mathrm{surv}}^{(N)}$ of
Eq.~\eqref{eq:zenoSurvival}, for any $N$. 

Away from the Zeno limit the
heralded state keeps one exact property: writing
Eq.~\eqref{eq:minusState} as
$\ket{\phi_-} \propto \sigma_x^S(\Id - \hat\Sigma_M)\hat X\ket{0}_S$
shows that it is an eigenstate of the memory parity $\hat\Sigma_M$
with eigenvalue $-1$, for any $N$, any spacing, and any $\delta$ with
$p_- > 0$. 

To find the memory parity, we measure each memory in the $X$ basis and multiply the $+1$ or $-1$ results. For the minus schedule outcome, \begin{equation} \boxed{\text{Measured value of }\hat{\Sigma}_M=-1} \label{eq:memory-parity} \end{equation} When the schedule readout gives the minus outcome, measuring the memories in the $X$ basis always gives a product of $-1$ in the ideal protocol.
Together with the complement relation in Eq.~\eqref{eq:antizeno}, this is a main exact result of the protocol. In other words, every ideal run with a minus schedule outcome has an odd number of memory results equal to $-1$, so their product is always $-1$. The corresponding fully dephased schedule gives positive and negative parity equally often. Measuring this parity provides an additional test of coherence between the schedules without relying on the system qubit's final state.

\begin{figure}[t]
\centering
\includegraphics[width=3.404in]{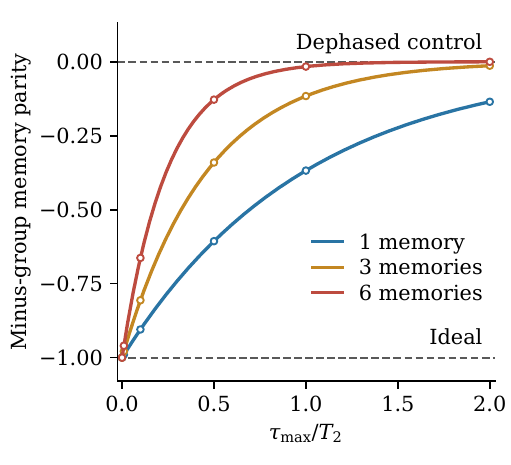}
\caption{Minus-selected memory parity versus storage time relative to coherence time. Curves give Eq.~\eqref{eq:parity-memory-decay} for $N=1,3,6$. Markers are direct density-matrix calculations. The horizontal references are the ideal coherent value $-1$ and the matched dephased-schedule value zero. We hold the timing fixed and vary $T_2$, using $T_1=T_\phi=3T_2/2$. The illustrative sequence has $\Omega/(2\pi)=50$ kHz, $\Omega T=3$, $\Omega\delta=\pi/2$, recording slots of $0.18\,\mu\mathrm{s}$, an offset slot of $5\,\mu\mathrm{s}$, and a final wait of $0.2\,\mu\mathrm{s}$. So $\tau_k=(N-k)(T/N+0.18\,\mu\mathrm{s})+5.2\,\mu\mathrm{s}$. Only post-recording memory noise is included. $Q$, $S$, gates, preparation, and readout are ideal. These are model expectations, not device measurements.}
\label{fig:parity-coherence}
\end{figure}

Fig.~\ref{fig:parity-coherence} shows how finite memory coherence attenuates this ideal parity signal under a specified storage-noise model. This exact parity constrains the complementary runs as well: because the unread ensemble has zero
memory parity overall, the plus-selected subset must compensate,
giving $\langle\hat\Sigma_M\rangle_+ = p_-/p_+$.

In the computational basis, $S$ in the minus-selected runs looks like the
flipped Zeno reference. The match stops there, because the heralded state of $S$ also
carries a $\sigma_y$ coherence that the dephased reference state lacks. This coherence can be checked by measuring $S$ in the $Y$ basis, giving another way to test the selected state alongside the probabilities discussed below.

Eq.~\eqref{eq:minusState} gives the structure of the heralded state, and its squared norm gives the probability $p_-$ of obtaining it, Eq.~\eqref{eq:ppm}. Eq.~\eqref{eq:antizeno} says that the antisymmetric outcome heralds a flip of $S$ on top of an otherwise standard Zeno trajectory, so its final-zero probability is the complement of the Zeno curve.  So at fixed $N$ and $T$, sweeping $\delta$ gives a flat $\ket{-}_Q$ branch, while the  $\ket{+}_Q$ branch varies with $\delta$, giving a clean experimental signature.

 In the superposed-schedule circuit itself, ignoring the readout outcome gives Eq.~\eqref{eq:marginal} rather than Eq.~\eqref{eq:zenoUnconditional}, and by Eq.~\eqref{eq:margLimit} that mixture saturates at $\mathcal{V}$ instead of at unity. The branch structure survives in the superposed circuit. For offsets with $0<\mathcal{V}<1$, the minus-selected probability approaches zero, while the plus-selected probability and the unread probability approach nonzero values. When $\mathcal{V}=0$, all three approach zero. When $\mathcal{V}=1$, the minus outcome does not occur. Sorting a single dataset by the outcome of $Q$ isolates a subset of runs whose final-zero probability is the complement of the Zeno curve.

The choice of $\delta$ determines both how often the $\ket{-}_Q$ outcome occurs and how far apart the two conditional probabilities lie. When $\delta\rightarrow0$, the two schedules become identical, so $Q$ is almost always found in $\ket{+}_Q$ and $p_-\rightarrow0$. In this limit, the $\ket{+}_Q$ branch reduces to the standard Zeno result. As $\Omega\delta$ increases from zero to $\pi$, the minus outcome becomes more likely, reaching $p_- = 1/2$ at  $\Omega\delta=\pi$. At that point, however, the two schedules become orthogonal, $p_+=p_-=\tfrac12$, and the difference between the $\ket{+}_Q$ and $\ket{-}_Q$ results disappears. Therefore, an intermediate value of $\delta$ gives the clearest experimental signature. 

For example, at $\Omega\delta=\pi/2$, $\mathcal{V}=\tfrac12$ and $p_-=\tfrac14$, so the inverted branch happens in around one quarter of the experimental runs. For $\Omega T=3$ and $N=8$, this gives $P_0^{(N,-)}\simeq0.219$ against $P_0^{(N,+)}\simeq0.520$, a difference of approximately $0.30$ between the two readout outcomes within one dataset. Ignoring the readout outcome in the same dataset gives $\overline{P_0}^{(N)}\simeq0.445$, while the $\delta=0$ reference run gives the standard Zeno value $P_0^{(N)}\simeq0.781$, as shown in Fig.~\ref{fig:main}.

\subsection{Classical Versus Quantum Measurement Schedules}
\label{sec:classical}

Classical uncertainty in measurement times has been considered previously~\cite{EgusquizaGaray2003}. In that setting, each realization has a definite schedule, even though the times vary randomly between realizations. This differs from the coherent superposition of schedules considered here.

Every prediction above follows from standard quantum mechanics applied to a circuit where the schedule qubit coherently selects one of the two measurement schedules. While the motivation for the circuit comes from the Page--Wootters construction, the predictions themselves do not depend on adopting that framework; they require only that $Q$ retains coherence. If $Q$ were instead in a classical mixture of the two schedules, the system statistics would look identical as long as $Q$ was
not read out.

To show this concretely, we compare three models of the same experiment. In the first, the measurement times are set by a classical controller that chooses $\mathcal{A}$ or $\mathcal{B}$ at random with equal probability. The final state of $S$ is then the mixture $\tfrac12(\rho_S^{\mathcal{A}}+\rho_S^{\mathcal{B}})$, there is no coherence of $Q$, and no measurement in the $\ket{\pm}_Q$ basis is defined. In the second, the schedule qubit exists but is dephased, so the global state is $\tfrac12(\ket{A}\!\bra{A}\otimes\rho^{\mathcal{A}}+\ket{B}\!\bra{B}\otimes\rho^{\mathcal{B}})$, giving $\mathcal{V}=0$ and $P_0^{(\pm)}=\tfrac12(P_0^{\mathcal{A}}+P_0^{\mathcal{B}})$ regardless of $\delta$. In the third, $Q$ is coherent and Eqs.~\eqref{eq:visibility} and \eqref{eq:antizeno} apply. 

All three models give the same unread mixture of the two schedules, as described by Eq.~\eqref{eq:marginal}. This mixture is not generally the standard Zeno reference obtained with zero offset. Coherence between the schedules allows interference in the readout of $Q$ and can change the system probabilities conditioned on that readout. It does not change the system's unread probability.

This also means that the usual Zeno suppression by itself cannot tell us if the schedule qubit retains coherence, since
a classical measurement schedule
produces the same result.

Coherently controlling when an operation happens has also been studied before. Henderson \textit{et al.}~\cite{Henderson2020} superposed two particle detectors switched on at different spacetime points and showed both that interference between the branches reveals field correlations neither branch shows alone, and that tracing out the control leaves an ordinary classical mixture. 

Our circuit reproduces the branch structure of a quantum switch~\cite{Chiribella2013,Procopio2015,Goswami2018}. Its two branches are $\hat U_S(\delta)\hat X$ and $\hat X\hat U_S(\delta)$, so the offset rotation and the recording sequence are applied in opposite orders. Superposing the start time of the sequence relative to a drive that is always on amounts to superposing the order of the sequence and one segment of that drive, and a similar link between superposed timing and superposed order appears in Ref.~\cite{Zych2019}. The idea of reading such a control qubit after it has steered an experiment goes back to the quantum delayed-choice proposal of Ionicioiu and Terno~\cite{IonicioiuTerno2011}. 

We implement the switch with two controlled rotations in a fixed-order circuit, in which the offset rotation appears once in each position, so no indefinite causal order is involved. This differs from single-use implementations of the switch, whose interpretation has been debated (see, e.g., Ref.~\cite{Oreshkov2019}), and none of our results depend on that question. 

The switch structure fixes what the readout of $Q$ measures. Its $\ket{\pm}_Q$ outcomes project onto the anticommutator and commutator of the two operations~\cite{Chiribella2012,Procopio2015}. In our repeated-measurement setting these become a visibility that does not depend on the number or spacing of the records and a minus branch that is the flipped Zeno reference, Eq.~\eqref{eq:minusState}.

\section{Proposed Experimental Test}
\label{sec:experiment}

The proposed experiment differs from earlier optical demonstrations of Page--Wootters clock correlations using entangled photon pairs~\cite{Moreva2014,Moreva2017}. Here, $Q$ does not serve as a clock, but instead controls the placement of an offset rotation. The circuit nonetheless has a stationary history representation, which we discuss in Appendix~\ref{app:embedding}.

The proposed circuit in Fig.~\ref{fig:circuit}(b)  uses $N+2$ processor qubits, consisting of one schedule qubit $Q$, one system qubit $S$, and $N$ memory qubits. The memory qubits are what let the records themselves be read out directly. Without them, the statistics of $Q$ and $S$ depend only on the dephasing channel that each record applies to $S$,  so any process with the same dephasing would reproduce $p_\pm$ and $P_0^{(N,\pm)}$. 

The circuit requires $N$ recording CNOTs and two controlled rotations. Each controlled rotation can be decomposed into two CNOTs and single-qubit gates, giving $N+4$ two-qubit gates in total. The routing overhead depends on the hardware connectivity: $Q$ couples only to $S$, while $S$ must interact with each memory qubit in sequence.

If we replaced the memory CNOTs with mid-circuit measurements of $S$, which also dephase $S$ in the computational basis, the resulting circuit would cross-check signatures (1), (2), and (3) below. However, that circuit would not be able to test the heralded memory state of Eq.~\eqref{eq:minusState}, which requires reading the memories, signature (4) below.

The experimental test has four signatures that test the models of Sec.~\ref{sec:classical}. For signatures (1)--(3), each run ends with measurements of $Q$ and $S$, with the memories left unread. Signature (4) also measures the memories in the $X$ basis. 

\emph{(1) Baseline.} With the recording CNOTs disabled and the controlled offset rotations retained, the two branches produce the same ideal final state. The fringe contrast of $Q$ should therefore be $\mathcal{V}=1$. Repeating this baseline at each value of $N$ provides a comparison for the recording-enabled circuit and helps characterize the effects of gate errors and decoherence.

\emph{(2) Visibility.} With the recording CNOTs enabled, the ideal prediction is $\mathcal{V}=\cos^2(\Omega\delta/2)$, independent of $N$ at fixed $\delta$. Measuring the visibility across different values of $N$ tests this prediction. Any observed dependence on $N$ would need to be assessed against gate errors, decoherence, and other departures from the ideal protocol. The baseline measurements provide a comparison, but do not by themselves separate these contributions.

\emph{(3) Complement relation.} As $N$ increases while $T$ and $\delta$ are kept fixed, the usual Zeno result makes $S$ increasingly likely to remain in $\ket{0}_S$. However, among the runs where $Q$ is found in $\ket{-}_Q$, the opposite happens, and $S$ becomes increasingly unlikely to remain in $\ket{0}_S$, as given by Eq.~\eqref{eq:antizeno}. Sorting the measurements of $S$ by the schedule outcome gives the $\ket{+}_Q$ and $\ket{-}_Q$ curves from one dataset. Both groups are collected under the same experimental conditions, but sorting does not automatically cancel preparation, gate, or readout errors. 

The standard Zeno reference $P_0^{(N)}$ must be measured in a separate run with $\delta=0$ and the recording CNOTs enabled. This differs from the coherence baseline in signature (1), where the recording CNOTs are disabled. Ignoring the schedule outcome in the superposed run instead gives Eq.~\eqref{eq:marginal}, which approaches $\mathcal{V}$ in the Zeno limit. Confirming that saturation is itself a check on the construction, since it ties the residual final-zero probability to the visibility measured in signature (2).

\emph{(4) Record parity.} In the runs where $Q$ is found in $\ket{-}_Q$, apply a Hadamard to every memory and measure it in the computational basis. For measured memory bits $m_1,\ldots,m_N\in\{0,1\}$, the sample average of $(-1)^{m_1+\cdots+m_N}$ estimates $\langle\hat\Sigma_M\rangle$, whose ideal value is exactly $-1$ for any $N$, any spacing, and any $\delta$ with $p_->0$. The plus-selected runs should give $\langle\hat\Sigma_M\rangle_+=p_-/p_+$, and a classical or dephased schedule gives zero in both. Unlike signatures (2) and (3), this one tests the records themselves. The parity is a product over $N$ memories, so its contrast falls with memory dephasing and readout error, and small $N$ is the natural regime for this test.

To quantify the loss of record parity, let $\tau_k$ be the storage time from the end of memory $k$'s recording gate to its final basis rotation. We consider independent relaxation and pure dephasing during this interval, with times $T_{1,k}$ and $T_{\phi,k}$, and define $T_{2,k}^{-1}=(2T_{1,k})^{-1}+T_{\phi,k}^{-1}$. The combined channel on each memory multiplies $\sigma_x^{M_k}$ by $e^{-\tau_k/T_{2,k}}$. With otherwise ideal operations and perfect readout, the minus-selected mean is therefore
\begin{equation}
\langle\hat\Sigma_M\rangle_-
=-\exp\!\left[-\sum_{k=1}^{N}\frac{\tau_k}{T_{2,k}}\right].
\label{eq:parity-memory-decay}
\end{equation}
A matched dephased-schedule control has zero mean under the same channels. Since each memory stops interacting after its record is written, its accumulated storage noise can be applied at the end without changing this result. 

This decay of memory parity does not imply a loss of schedule visibility. The post-recording memory channels considered here leave the reduced state of $Q$ unchanged.

Figure~\ref{fig:parity-coherence} compares these limits for equal memory coherence times $T_2$. Here $T_2$ is the Ramsey coherence time. The horizontal variable is $\tau_{\max}/T_2$, where $\tau_{\max}=\max_k\tau_k$ is the longest storage interval, rather than the full circuit duration. The parity approaches $-1$ when the accumulated exposure $\sum_k\tau_k/T_2$ is small and approaches zero as that exposure grows. So a sequence short compared with each memory's coherence time can still lose appreciable joint parity when many memories contribute. 

Finite coherence does not necessarily erase the contrast with the dephased control, but that contrast alone does not certify the full state or exclude every alternative preparation. The accompanying memory-parity notebook~\cite{SeidelParityNotebook2026} provides the calculation and data underlying Fig.~\ref{fig:parity-coherence} and checks Eq.~\eqref{eq:parity-memory-decay} against density-matrix evolution. It also separately quantifies the effects of memory readout errors and schedule-readout misclassification under the noise assumptions stated here.

Readout loss is separate from the storage effect shown here. Independent symmetric memory-bit errors $e_k$ add the factor $\prod_k(1-2e_k)$, while schedule-readout mistakes mix the plus and minus groups and change their conditional normalization. Infinite memory coherence does not recover $-1$ in the presence of readout errors. Asymmetric assignment errors can also shift the dephased control away from zero, so an experimental comparison needs a matched control and calibrated readout.

 \par \textit{Error budget.} Except for the specified memory channels in Fig.~\ref{fig:parity-coherence}, the predictions above assume ideal preparation, drive rotations, recording gates, and readout. In a hardware implementation, dephasing of $Q$, relaxation and dephasing of $S$, imperfect memory preparation, gate errors, and unwanted couplings can affect the measured statistics. Their effects depend on when they occur within the sequence.

Two systematic errors need particular care. First, a static $ZZ$ coupling between $Q$ and $S$ acts during the sequence as a rotation of $S$ about $z$ whose sign depends on the branch. The sequence is then no longer identical in the two branches, which is the assumption behind both Eq.~\eqref{eq:visibility} and Eq.~\eqref{eq:antizeno}. In a simulation with a phase $\phi=0.1$~rad per recording step, applied after each drive rotation as $e^{-i\phi\,\sigma_z^Q\sigma_z^S/2}$, $\mathcal{V}$ falls from $0.50$ to $0.38$ between $N=1$ and $N=10$, and $P_0^{(N,-)}$ exceeds Eq.~\eqref{eq:antizeno} by $0.17$ at $N=10$. This is a spurious $N$ dependence of exactly the kind signature (2) is meant to exclude. Because the sequence is not controlled by $Q$, a $\pi$ pulse on $Q$ at the midpoint of the sequence and a second one at its end leave the ideal protocol unchanged. In the same simulation, these gates reduce the error in $P_0^{(N,-)}$ below $5\times10^{-3}$ for all $N\geq2$ and the drift in $\mathcal{V}$ at $N=10$ to $0.015$. 

Second, a readout of $Q$ in a single basis cannot separate a loss of visibility from a rotation of the relative phase between the branches, which ac Stark shifts, residual $ZZ$, or miscalibrated controlled rotations can all produce. Scanning the phase of the final gate on $Q$ and fitting the fringe separates the two. Dephasing of $Q$ also moves weight from the plus outcome into the rarer minus outcome and biases $P_0^{(N,-)}$ toward the unread value. At $N=8$, a residual coherence factor of $0.9$ for $Q$ raises it from $0.219$ to $0.260$. The script that produces the ZZ, echo, and dephasing numbers above is in the repository of Ref.~\cite{SeidelParityNotebook2026}. The record-disabled baseline, whose ideal visibility is one, measures this factor directly, so it can be used to correct the conditional probabilities.

The CNOT-disabled baseline can help characterize the achievable schedule contrast. For a useful comparison, the baseline's duration and idle periods should match the recording-enabled circuit. However, removing the recording gates also removes their errors, so the baseline cannot automatically correct the full recording-enabled experiment.

Appendix~\ref{app:visibility} shows that dephasing, detuning, thermal preparation errors in $S$, and imperfect intermediate recordings do not change the schedule visibility. This result assumes a transverse offset rotation, no $Q$-dependent dynamics during the recording sequence, and a final record that is made after the modeled noise. The conditional system probabilities also require an error analysis, including preparation and readout errors. These checks can use measurements of $Q$ and $S$ without reconstructing the full joint state.

\section{Discussion and Conclusion}
\label{sec:discussion}
\subsection{Relational Interpretation of the Protocol}
\label{sec:stationarity}

The repeated-measurement sequence described above can be incorporated into a stationary Page--Wootters history state,
\begin{equation}
\ket{\psi_U^{(N)}}=\int dt\;\ket{t}_C\otimes\ket{\psi_{SM}^{(N)}(t)},
\label{eq:zenoHistoryState}
\end{equation}
where $t$ is the clock reading labeling points along the history, while $N$ and the measurement spacing $\Delta t=T/N$ are fixed for the protocol being represented. The state $\ket{\psi_{SM}^{(N)}(t)}$ contains the corresponding system--memory state at clock reading $t$. With the appropriate clock-dependent constraint, this history can be represented within the same stationary framework discussed in Appendix~\ref{sec:assumptions}. However, this stationary representation is not specific to the quantum Zeno effect. More generally, any unitary history can be encoded as correlations between clock states and the corresponding states of the rest of the system~\cite{Diaz2025}.

Stationarity by itself does not give a distinctive signature of relational Zeno dynamics. Writing the Zeno evolution as a stationary Page–Wootters history does not change its predicted probabilities, and observing standard Zeno suppression relative to $C$ would reproduce the standard quantum Zeno effect~\cite{Itano1990} regardless of whether the underlying description is stationary or not. The new ingredient is treating the measurement schedule as a quantum degree of freedom. 

A separate consequence of having physical memory qubits is that the protocol allows a limited comparison with Wigner’s-friend scenarios~\cite{Baumann2021}. We can regard the memories as an observer's records and ask how the system statistics depend on the basis used to read the schedule qubit. 

Reading $Q$ in the $\{\ket{A},\ket{B}\}$ basis distinguishes the schedules, while reading it in the $\{\ket{+},\ket{-}\}$ basis gives the conditional probabilities derived above. This choice changes how the system outcomes are grouped, but it does not change their unread distribution; it changes what can be said about the records. Conditioned on the minus outcome and in the Zeno limit, the records are left in the superposition $(\ket{0\cdots0}_M-\ket{1\cdots1}_M)/\sqrt2$ of every record reading zero and every record reading one, Eq.~\eqref{eq:minusState}. For any $N$ the records are left in an eigenstate of their joint $X$-basis parity with eigenvalue $-1$, which signature (4) tests and which a classical record of either schedule cannot show.

The finite-coherence comparison in Fig.~\ref{fig:parity-coherence} shows how the memory-parity contrast with the matched dephased control weakens as the memories store their records. \par \subsection{Conclusions and Outlook}
We have studied a repeated-measurement circuit in which a schedule qubit coherently controls whether an additional system rotation occurs before or after a recording sequence. For the ideal protocol with recording interactions present, the schedule visibility is $\mathcal{V}=\cos^2(\Omega\delta/2)$, Eq.~\eqref{eq:visibility}, independent of the number of recordings and their spacing. Removing the recording interactions restores full visibility.

Selecting the minus outcome of the schedule readout gives $P_0^{(N,-)}(T)=1-P_0^{(N)}(T)$, Eq.~\eqref{eq:antizeno}, where $P_0^{(N)}(T)$ is the final-zero probability in a separate zero-offset reference run. This relation applies whenever the minus outcome has nonzero probability. In the Zeno limit, the selected probability approaches zero while the reference approaches one. Ignoring the schedule outcome instead gives a mixture whose final-zero probability approaches $\mathcal{V}$.

These predictions can be tested by measuring the schedule qubit and the system; reading the memory parity tests the recorded superposition directly. An accompanying notebook provides preliminary estimates of the gate counts and routing overhead for implementing the proposed protocol on two different types of cloud-based quantum processors from IBM, Heron and Nighthawk~\cite{SussmanSeidel2026ZenoFeasibility}.

Several extensions of the protocol are possible. A higher-dimensional schedule register could place more than two measurement schedules in superposition, including schedules with different numbers of measurements $N$. This would allow the measurement frequency itself to be treated as a quantum variable. 

We could also allow the individual measurement times to shift independently rather than moving the whole measurement sequence by the same offset $\delta$. In that case, the visibility would likely depend on $N$, because the intermediate parts of the two schedules would not match. Measuring how the visibility changes could then show how much information about the measurement schedule is stored in each memory record. 

Finally, the memory register could be coherently erased before the final readout. If the loss of coherence in $S$ comes from quantum information stored in the memories, removing that information should restore the coherence in $S$, providing an additional experimental check of the mechanism.

\begin{acknowledgments}
O.S. would like to thank Matthew Skrzypczyk for helpful discussions about whether the ricochet property could account for the effects described here, and Daniel Baxter for a careful reading of the manuscript and helpful comments on its presentation and conceptual clarity. O.S. used Anthropic's Claude (Opus 5) to check the derivations in Sec.~\ref{sec:superposed}, to run the numerical checks behind the error budget in Sec.~\ref{sec:experiment}, and to draft the feasibility notebook of Ref.~\cite{SussmanSeidel2026ZenoFeasibility}. The authors verified every result independently and take full responsibility for the content. O.S. was supported by a Graduate Instrumentation Research Award (GIRA) through the Coordinating Panel for Advanced Detectors (CPAD) as funded by the DOE Office of High Energy Physics. This work was produced by Fermi Forward Discovery Group, LLC under Contract No. 89243024CSC000002 with the U.S. Department of Energy, Office of Science, Office of High Energy Physics. Publisher acknowledges the U.S. Government license to provide public access under the DOE Public Access Plan (http://energy.gov/downloads/doe-public-access-plan).
\end{acknowledgments}

\appendix
\section{The Page--Wootters Framework for Relational Time}
This appendix reviews the Page--Wootters background used to interpret the circuit in the main text. We introduce the stationary global state, derive the conditional Schr\"odinger evolution, and discuss the assumptions of this description. Appendix~\ref{app:embedding} gives the finite history construction for our protocol. Throughout, we use units where $\hbar=1$.

\subsection{A Static Universe}

In standard quantum mechanics, a state $\ket{\psi(t)}$ evolves with
respect to a time $t$ that is an external parameter, not an observable. Time therefore is treated differently from dynamical variables such as position and momentum. Those are promoted to self-adjoint operators on the Hilbert space, while $t$ stays a scalar label alongside parameters such as masses and coupling constants. Pauli's argument is the usual reason given for this, as a self-adjoint operator canonically conjugate to the Hamiltonian would generate energy translations to arbitrarily negative values, which is incompatible with a spectrum bounded from below~\cite{Pauli1980}. In the relational approach, one quantum subsystem serves as a clock for another. The clock's time observable is conjugate to the clock Hamiltonian, while the Hamiltonian of the system whose evolution is being described acts on a different subsystem~\cite{LeonMaccone2017}.

The Page--Wootters construction promotes the reading of a clock subsystem, rather than time itself, to a quantum variable~\cite{PageWootters1983}. To see why treating a clock as a quantum subsystem is natural, consider the case where the system in question is the whole universe. There is then no outside clock to refer to. If we represent the quantum state of the whole universe as $\ket{\psi_U}$, we can expand its state in the energy basis to be in a superposition of energy eigenstates, so $\ket{\psi_U} = \sum_n c_n \ket{E_n}_U$, as we can do for any arbitrary state. 

Standard quantum mechanics would say that this state time-evolves according to the dynamics set by a Hamiltonian operator $\hat{H}$ (we use $\hat{H}_U$ in this paper to denote the Hamiltonian governing the dynamics for the whole universe), so that $\ket{\psi_U}$ acquires time dependence, $\ket{\psi_U(t)}$.

In this standard description, however, the parameter $t$ is supplied externally rather than represented by a degree of freedom within the universe itself. For a closed universe there is, by definition, no external physical system that can provide such a time reference. This motivates the Page--Wootters picture, in which the global state stays stationary while time and ordinary dynamics emerge internally from correlations between quantum subsystems~\cite{PageWootters1983,Wootters1984}.

The idea of a stationary universe is not only introduced for convenience. It also appears naturally in some formulations of quantum gravity, where physical states satisfy the Wheeler--DeWitt constraint. The stationary-state constraint then becomes~\cite{DeWitt1967} 
\begin{equation}
\hat H_U\ket{\psi_U}=0
\end{equation}
In this setting, the equation selects the allowed states of the whole universe. Applying $\hat H_U$ to such a state gives zero. Writing the condition this way makes the mathematics clearer by distinguishing the constraint on the whole universe from the energies of its individual parts.

More generally, the Page--Wootters construction only requires the global state to be stationary. We can therefore take it to be a definite total-energy eigenstate,
\begin{equation}
\hat H_U\ket{\psi_U}=E\ket{\psi_U},
\end{equation}
while the subsystems within it can still exhibit nontrivial dynamics relative to one another~\cite{PageWootters1983,Wootters1984}. Since an overall energy offset does not affect the relational dynamics, we will later choose the zero of energy so that $E=0$ for convenience. The construction has since been reformulated to address the ambiguities originally raised against it and to put it in a form suitable for concrete applications~\cite{MarlettoVedral2017}.

Now that we have established that the universe is in a definite total-energy eigenstate, we can consider the case where this energy eigenvalue $E$ is highly degenerate, giving many distinct basis states that all correspond to the same total energy $E$. We label these different states $\ket{E_a}_U$, $\ket{E_b}_U$, $\ket{E_c}_U$, and so on, where the subscripts $a,b,c,\dots$ distinguish different states with the same energy $E$.

The state of the universe can therefore be written as
\begin{equation}
\ket{\psi_U}=x\ket{E_a}_U+y\ket{E_b}_U+z\ket{E_c}_U+\dots .
\end{equation}

The state is normalized, so the squared magnitudes of all of the amplitudes add up to 1,
\begin{equation}
|x|^2 + |y|^2 + |z|^2 + \dots = 1.
\end{equation}

Because each of these different basis states has the same total energy $E$, we can write
\begin{equation}
\begin{array}{rcl}
\hat{H}_U\ket{E_a}_U & = & E\ket{E_a}_U,\\
\hat{H}_U\ket{E_b}_U & = & E\ket{E_b}_U,\\
\hat{H}_U\ket{E_c}_U & = & E\ket{E_c}_U,\\
& \vdots &
\end{array}
\end{equation}

Therefore, their superposition is also an eigenstate of the total Hamiltonian with the
same definite energy,
\begin{equation}
\hat{H}_U\ket{\psi_U}=E\ket{\psi_U}
\end{equation}

\noindent The superposition has a definite total energy even though it contains many different basis states.

Because $\ket{\psi_U}$ is a single total-energy eigenstate, applying the unitary generated by $\hat H_U$ changes the state only by an overall phase,
\begin{equation}
e^{-iEs}\ket{\psi_U},
\end{equation}
where $s$ is simply the parameter labeling this unitary transformation and is not yet the internal clock reading introduced below. This overall phase is not observable because it does not change any measurement outcomes. 

The Page--Wootters construction shows that nontrivial dynamics can nevertheless appear within this stationary global state when one subsystem is described relative to another. We make this explicit in Secs.~\ref{sec:clockreadings} and~\ref{sec:schrodinger-recovery}, where one subsystem is designated as a clock and the Schr\"odinger evolution of the remaining degrees of freedom is recovered relative to its readings.

In this picture, time does not have to be set by a clock external to the universe. Instead, time can be the parameter labeling correlations between two quantum subsystems, one designated as the `clock' and the other comprising the remaining degrees of freedom, which we call the `rest'.

\subsection{Splitting the Frozen Universe into a Clock and a Rest}

Consider the simplest version of a universe divided into two subsystems, a clock $C$ and the remaining subsystem $R$, which we refer to as the `rest'. The Hilbert space of the whole universe can then be split into the combined Hilbert spaces of these two subsystems,
\begin{equation}
\mathcal{H}_U = \mathcal{H}_C \otimes \mathcal{H}_R .
\end{equation}

The Page--Wootters picture assumes that the two parts do not interact, so the total Hamiltonian of the universe is
\begin{equation}
\hat{H}_U=\hat{H}_C \otimes \mathbb{I}_R+\mathbb{I}_C \otimes \hat{H}_R ,
\label{eq:Hsum}
\end{equation}

where $\hat{H}_C$ acts only on the clock and $\hat{H}_R$ acts only on the rest. For simplicity, we will suppress the identity operators moving forward, and write this as
\begin{equation}
\hat{H}_U = \hat{H}_C + \hat{H}_R .
\label{eq:hamiltonian}
\end{equation}

The frozen state of the universe is $\ket{\psi_U}$, the stationary definite-total-energy state introduced in the previous section. It therefore satisfies
\begin{equation}
\hat{H}_U\ket{\psi_U} = E\ket{\psi_U}.
\end{equation}

As noted above, we choose the zero of energy so that $E=0$. This gives the same zero-eigenvalue form as the Wheeler--DeWitt constraint discussed earlier, although here it is simply a convenient choice of energy reference within the Page--Wootters construction. The stationary-state constraint then becomes
\begin{equation}
\boxed{\hat{H}_U\ket{\psi_U}=0}
\label{eq:constraint}
\end{equation}

Eq.~\eqref{eq:constraint} is a constraint rather than an equation of motion. By Stone's theorem, because $\hat H_U$ is self-adjoint, it generates a one-parameter group of unitaries~\cite{Stone1930,HohnSmithLock2021}. With our choice $E=0$, the constraint makes $\ket{\psi_U}$ exactly invariant under this whole group,
\begin{equation}
e^{-i\hat{H}_U s}\ket{\psi_U}=\ket{\psi_U} \quad\text{for all } s.
\end{equation}

Here $s$ is a group parameter rather than a time, so no external time is needed to state this, and no time can be attached to the universe as a whole because nothing in the state changes under the group action. This is the structure of the Hamiltonian constraint in quantum gravity, where the constraint generates reparameterizations rather than time evolution~\cite{DeWitt1967,HohnSmithLock2021}. If we had instead kept a nonzero eigenvalue $E$, the same group action would give only the global phase $e^{-iEs}\ket{\psi_U}$ rather than exact equality with $\ket{\psi_U}$. Since this phase has no observable effect, the physical stationarity is unchanged.

\subsection{Reading the Rest Relative to the Clock}
\label{sec:clockreadings}
We now need to define what it means for the subsystem $C$ to act as a clock. This clock is not evolving with respect to some external time, and the global state $\ket{\psi_U}$ is still stationary. Instead, we define a family of possible states of the clock and label these states by a number $t$.

We choose a reference clock state $\ket{0}_C$. We then define the family of clock states to be
\begin{equation}
\ket{t}_C= e^{-i\hat{H}_C t}\ket{0}_C, \qquad t\in\mathbb{R}.
\label{eq:clockstates}
\end{equation}
At this point, $t$ should not be interpreted as the external time parameter used in standard quantum mechanics. It is simply a label that distinguishes different possible states, or readings, of the quantum clock.

These clock states are defined so that their labels behave like the readings of a quantum clock. For example, say the clock is in the state labeled $\ket{t}_C$. Applying the clock translation operator $e^{-i\hat{H}_C s}$ gives
\begin{align}
e^{-i\hat{H}_C s}\ket{t}_C&=e^{-i\hat{H}_C s}e^{-i\hat{H}_C t}\ket{0}_C \\
&=e^{-i\hat{H}_C(t+s)}\ket{0}_C \\
&=\ket{t+s}_C.
\label{eq:clocktranslation}
\end{align}

Here, $s$ is just a parameter labeling a translation between clock states, rather than being an external time through which the universe evolves. 

The whole universe remains in the single stationary state $\ket{\psi_U}$, and the different possible clock readings $\ket{t}_C$ label the corresponding states of the rest of the universe. At this stage, $t$ is only a clock-state label. The next section shows that the corresponding state of the rest obeys the Schr\"odinger equation with respect to $t$, which is what gives $t$ the interpretation of an internal time.

\begin{equation}
\ket{\psi_R(t)}={}_C\!\braket{t|\psi_U}.
\label{eq:relstate}
\end{equation}

In this expression, we project the clock part of the stationary global state $\ket{\psi_U}$ onto the clock state $\ket{t}_C$. What remains after this projection is a state living only in the Hilbert space $\mathcal{H}_R$ of the `rest' subsystem.

Different clock readings can therefore be correlated with different states of the `rest' subsystem,
\begin{equation}
\ket{t_1}_C\longleftrightarrow\ket{\psi_R(t_1)}, \qquad \ket{t_2}_C\longleftrightarrow\ket{\psi_R(t_2)}, \quad \dots
\end{equation}

even though the global state $\ket{\psi_U}$ itself never changes. Here, the sequence of states of the `rest' subsystem is held in the correlations between the clock and the rest within the single stationary global state.

The next section shows that, as the clock label $t$ changes, the corresponding state $\ket{\psi_R(t)}$ of the rest obeys the standard Schr\"odinger equation generated by $\hat{H}_R$. For this reason, the parameter $t$ that initially served only as a label for the clock states can hereafter be used as an internal time for the rest of the universe.

\subsection{Recovering Schr\"odinger Evolution}
\label{sec:schrodinger-recovery}
We can now show that the state of the rest associated with the clock reading $t$, $\ket{\psi_R(t)}$, follows the standard Schr\"odinger equation with respect to the internal clock reading $t$. Keeping in mind Eq.~\eqref{eq:relstate}, we differentiate this state with respect to the clock label $t$. From the definition of the clock states in Eq.~\eqref{eq:clockstates}, the corresponding bra is
\begin{equation}
{}_C\!\bra{t} = {}_C\!\bra{0}\,e^{+i\hat{H}_C t}.
\end{equation}
Therefore,
\begin{equation}
\partial_t\,{}_C\!\bra{t} = i\,{}_C\!\bra{t}\hat{H}_C.
\end{equation}

Using this result in the definition of $\ket{\psi_R(t)}$, we get
\begin{align}
i\,\partial_t\ket{\psi_R(t)}
&= i\,\partial_t\left({}_C\!\braket{t|\psi_U}\right) \nonumber\\
&= i\left(i\,{}_C\!\bra{t}\hat{H}_C\right)\ket{\psi_U} \nonumber\\
&= -\,{}_C\!\bra{t}\hat{H}_C\ket{\psi_U}.
\label{eq:step1}
\end{align}

Now we use the stationary-state constraint for the whole universe. From Eqs.~\eqref{eq:constraint} and \eqref{eq:hamiltonian} we have
\begin{equation}
\left(\hat{H}_C+\hat{H}_R\right)\ket{\psi_U}=0.
\end{equation}
Rearranging gives
\begin{equation}
\hat{H}_C\ket{\psi_U}=-\hat{H}_R\ket{\psi_U}.
\label{eq:Htrade}
\end{equation}
Substituting Eq.~\eqref{eq:Htrade} into Eq.~\eqref{eq:step1} gives
\begin{align}
i\,\partial_t\ket{\psi_R(t)}
&= -\,{}_C\!\bra{t}\left(-\hat{H}_R\right)\ket{\psi_U} \nonumber\\
&= {}_C\!\bra{t}\hat{H}_R\ket{\psi_U}.
\end{align}

Now, since $\hat{H}_R$ acts only on the rest subsystem and not on the clock, it can be moved outside the clock bra,
\begin{equation}
{}_C\!\bra{t}\hat{H}_R\ket{\psi_U}=\hat{H}_R\left({}_C\!\braket{t|\psi_U}\right).
\end{equation}
Using the definition $\ket{\psi_R(t)}={}_C\!\braket{t|\psi_U}$ then gives
\begin{equation}
i\,\partial_t\ket{\psi_R(t)}=\hat{H}_R\ket{\psi_R(t)}
\label{eq:schrodinger}
\end{equation}

This is exactly the Schr\"odinger equation for the rest subsystem, except now the parameter $t$ is the reading of the internal quantum clock rather than an external, classical time parameter. The global state $\ket{\psi_U}$ remains stationary, but the state of the `rest' subsystem associated with different clock readings changes according to standard Schr\"odinger dynamics. In this sense, the time evolution of the rest emerges from the correlations between the clock and the rest that are already contained in the frozen global state.

\subsection{An Explicit Frozen Universe State}

We now write an explicit form of the stationary state $\ket{\psi_U}$. Suppose the clock is idealized so that its different readings are perfectly distinguishable, so that
\begin{equation}
{}_C\!\braket{t'|t}=\delta(t'-t).
\end{equation}
We can now construct the global state
\begin{equation}
\begin{aligned}
\ket{\psi_U} 
&=\int_{-\infty}^{\infty}dt\;\ket{t}_C\otimes\ket{\psi_R(t)},\\
\ket{\psi_R(t)}
&=e^{-i\hat H_R t}\ket{\psi_R(0)}.
\end{aligned}
\label{eq:history}
\end{equation}

This state has every clock reading $\ket{t}_C$ correlated with the corresponding state $\ket{\psi_R(t)}$ of the rest. Rather than imagining that the universe moves from one of these terms to the next, all of the clock-rest correlations are held together in the single global state $\ket{\psi_U}$. As a result, the whole history of the rest is contained in one stationary state, with no external clock singling out any particular moment as the present.

The continuous history in Eq.~\eqref{eq:history} is a formal, non-normalizable state: with perfectly distinguishable clock readings and a normalized rest state at every reading, its norm contains the divergent integral $\int_{\mathbb{R}}dt$. It is therefore understood as a generalized state, with the constraint interpreted in the distributional sense, as in Ref.~\cite{GLM2015}.

In the clock-reading representation, $\hat H_C$ acts as $-i\partial_t$. Using the relational Schr\"odinger equation, we obtain
\begin{equation}
\begin{aligned}
{}_C\!\bra{t}\hat H_U\ket{\psi_U} &= \left(-i\partial_t+\hat H_R\right)\ket{\psi_R(t)} \\
&= 0.
\end{aligned}
\label{eq:checkstatic}
\end{equation}
So the history satisfies the stationary constraint in this generalized sense. This argument does not require assigning vanishing values to the clock-state kets at $t=\pm\infty$.

For our circuit, Appendix~\ref{app:embedding} gives a finite, normalized history state whose stationarity can be checked directly, without the continuous-clock idealization.

The whole state $\ket{\psi_U}$ is stationary, while projecting the clock onto a particular reading gives
\begin{equation}
{}_C\!\braket{t|\psi_U}\propto\ket{\psi_R(t)}.
\end{equation}
After normalization, this is exactly the state of the rest associated with the clock reading $t$. This does not introduce a new physical result beyond the relational dynamics derived above, but rather gives an explicit stationary state that contains all of those clock--rest correlations at once and verifies directly that the Page--Wootters construction can encode the full history in a single frozen global state.

The continuous clock used above is an idealized model. A perfectly continuous quantum clock with exactly distinguishable time states runs into the usual Pauli objection, because it would need an unbounded clock Hamiltonian, as described in~\cite{LeonMaccone2017}. A realistic clock has a finite number of states, so its different time readings are only approximately distinguishable~\cite{HohnSmithLock2021}. For the protocol considered here, Appendix~\ref{app:embedding} instead uses a finite register with one orthogonal clock label per circuit stage to construct a stationary history. This register belongs to the mathematical representation and is not an additional register required by the proposed experiment.

\subsection{The Relational Choice of Clock}

So far we have chosen the subsystem $C$ to act as the clock and $R$ to act as the rest. However, there is nothing fundamentally special about the label $C$ or about the particular subsystem initially chosen to serve as the clock. This freedom in the choice of clock, and of which clock observable is used, was one of the main objections raised against the Page--Wootters construction, since different choices appeared to give different dynamics for the rest. Marletto and Vedral~\cite{MarlettoVedral2017} showed that no such ambiguity arises in unitary quantum theory, provided the clock and clock observable are chosen so that Schr\"odinger-type dynamics is generated on the rest, and reformulated the model in a form intended to support concrete applications.

If the subsystem $R$ itself has the properties required to function as a suitable clock, we can instead use states of $R$ as our reference and describe the state of $C$ relative to them.

Let's suppose $R$ has a family of possible clock states
\begin{equation}
\ket{r}_R=e^{-i\hat H_Rr}\ket{0}_R,
\end{equation}
where $r$ labels the different readings of $R$ when it's being used as the clock. We can instead use $R$ as the reference and define the corresponding state of $C$ when $R$ has the reading $r$ as
\begin{equation}
\ket{\psi_C(r)}={}_R\!\braket{r|\psi_U}.
\end{equation}

The same stationary-state constraint,
\begin{equation}
\left(\hat H_C+\hat H_R\right)\ket{\psi_U}=0,
\end{equation}
now gives the same derivation with the roles of $C$ and $R$ exchanged. This gives
\begin{equation}
i\,\partial_r\ket{\psi_C(r)}=\hat H_C\ket{\psi_C(r)}.
\label{eq:schrodingerC}
\end{equation}

Therefore, when $C$ is used as the clock, the rest subsystem evolves according to $\hat H_R$, while if $R$ is itself capable of serving as a good clock, $C$ can instead be described as evolving according to $\hat H_C$ relative to the readings of $R$. In both descriptions, the global state $\ket{\psi_U}$ remains stationary. The evolution is due to the relation between the two subsystems, rather than to an evolution of the combined state with respect to an external classical time.

\subsection{Assumptions and Limits of the Relational-Time Construction}
\label{sec:assumptions}

The following assumptions clarify how this background relates to the protocol in the main text.

\emph{(i) The clock--rest correlation is part of the frozen state, not a process happening in classical time.} The correlations between the clock states $\ket{t}_C$ and the corresponding rest states $\ket{\psi_R(t)}$ are already contained in the single stationary state $\ket{\psi_U}$. The clock does not physically evolve through an external time and become entangled with the rest one step at a time. Instead, the different clock readings and their corresponding rest states are all encoded together in the fixed global state, and the apparent evolution only appears when we describe the rest relative to different clock readings.

\emph{(ii) The clock and the rest have been assumed not to interact.} In the simple Page--Wootters picture used above, the total Hamiltonian is $\hat H_U=\hat H_C+\hat H_R$, so the clock and the rest evolve without directly affecting one another. This is why the state of the rest relative to the clock reading $t$, $\ket{\psi_R(t)}$, follows the standard Schr\"odinger equation. If we instead include a coupling between them,
\begin{equation}
\hat H_U=\hat H_C+\hat H_R+\hat H_{CR},
\end{equation}
then the simple derivation changes, and the conditional dynamics can include interaction-dependent terms~\cite{SmithAhmadi2019}.

In our circuit, the schedule qubit $Q$ belongs to the rest subsystem together with $S$, the memories, and the readout record. Its controlled rotations are interactions within that subsystem. They are not themselves interactions between the reference clock $C$ and the rest.

Appendix~\ref{app:embedding} uses a separate finite clock register $\mathcal{T}$ to represent the gate sequence as a stationary history. Its propagation constraint couples neighboring clock labels to the corresponding gates on the rest. This construction explicitly includes the gate sequence and differs from the noninteracting clock--rest model used in the derivation above. The controlled schedule operations and the history-clock coupling therefore have separate roles.

\emph{(iii) Sequential measurements require care in the Page--Wootters picture.} The original Page--Wootters proposal~\cite{PageWootters1983} raised questions about how to consistently describe correlations involving measurements made at more than one clock reading. Later work developed ways to treat these sequential measurements consistently and recover the expected measurement statistics~\cite{GLM2015}, and multi-time correlations of this kind have been probed experimentally in an optical Page--Wootters setting~\cite{Moreva2017}. These later developments are important for any extension of the framework that involves repeated measurements or measurement-induced dynamics.

\section{The Protocol as a Stationary History State}
\label{app:embedding}

The circuit and its final readout can be represented by a single stationary Page–Wootters history state. Measurement outcomes are encoded as correlations within this state, following Giovannetti, Lloyd and Maccone~\cite{GLM2015}. The construction pairs the state after each gate with a distinct clock reading.

We first list the gates of the circuit in the order they act, as $U_1,\dots,U_K$. $U_1$ is the rotation $R_x(\Omega\delta)$ on $S$ controlled on $\ket{B}_Q$. $U_2,\dots,U_{2N+1}$ alternate the drive rotation $R_x(\Omega\Delta t)$ on $S$ and a CNOT from $S$ to the next memory qubit. $U_{2N+2}$ is $R_x(\Omega\delta)$ on $S$ controlled on $\ket{A}_Q$. $U_{2N+3}$ is the Hadamard on $Q$ and $U_{2N+4}$ is the CNOT from $Q$ to $W$, so $K=2N+4$. 

Let $\ket{\psi_0}$ be the initial state with $Q$ in $\ket{\chi}_Q$ and $S$, all $M_k$, and $W$ in $\ket{0}$, and let $\ket{\psi_k}=U_k\cdots U_1\ket{\psi_0}$ be the state after $k$ gates.

The clock is a register $\mathcal{T}$ with $K+1$ readings $\ket{k}_{\mathcal{T}}$, one for each stage of the circuit, so that reading $k$ means $k$ gates have happened. We denote this finite clock by $\mathcal{T}$ to distinguish it from the continuous Page--Wootters clock $C$ used in Sec.~\ref{sec:clockreadings}. The history state is
\begin{equation}
\ket{\Psi_U}=\frac{1}{\sqrt{K+1}}\sum_{k=0}^{K}\ket{k}_{\mathcal{T}}\otimes\ket{\psi_k}.
\label{eq:discreteHistory}
\end{equation}
Each clock reading is paired with the snapshot of the qubits at that stage, and all of them are held together in one state. This is the same construction as Eq.~\eqref{eq:zenoHistoryState}, with the continuous clock replaced by one that counts gates. Nothing evolves in this state. It is a zero-energy eigenstate of the constraint operator
\begin{equation}
\begin{split}
\hat H_{\mathrm{clock}}
&=\frac12\sum_{k=0}^{K-1}\Bigl[
\\
&\quad
\bigl(\ket{k}\!\bra{k}_{\mathcal{T}}
+\ket{k+1}\!\bra{k+1}_{\mathcal{T}}\bigr)
\otimes\Id
\\
&\quad
-\ket{k+1}\!\bra{k}_{\mathcal{T}}
\otimes U_{k+1}
\\
&\quad
-\ket{k}\!\bra{k+1}_{\mathcal{T}}
\otimes U_{k+1}^{\dagger}
\Bigr].
\end{split}
\label{eq:feynmanConstraint}
\end{equation}
which one can check term by term, since each bracket sends the pair of neighboring stages $\ket{k}_{\mathcal{T}}\ket{\psi_k}+\ket{k{+}1}_{\mathcal{T}}\ket{\psi_{k+1}}$ to zero. 

This operator is the propagation term of Kitaev's circuit-to-Hamiltonian construction~\cite{Kitaev2002}, a form of Feynman's clock~\cite{Feynman1986} also used for this purpose in Ref.~\cite{Diaz2025}, and it plays the role of the constraint in Eq.~\eqref{eq:constraint}. 
Its zero-energy eigenspace is degenerate, because every initial state $\ket{\psi_0}$ generates a history of this form. The initial state is therefore an input, just as the gate sequence is. It can be fixed by hand, or by adding a term that penalizes any history whose component at $\mathcal{T}=0$ differs from $\ket{\psi_0}$, as in Kitaev's construction~\cite{Kitaev2002}, which makes $\ket{\Psi_U}$ the unique zero-energy state. 

Here the clock and the rest do interact, because which gate acts depends on what the clock reads. This is the situation anticipated in Sec.~\ref{sec:assumptions}(ii). The state $\ket{\Psi_U}$ itself never changes. The whole experiment, from the first gate to the readout of $Q$, is stored in it as correlations between what the clock reads and what $Q$, $S$, $M$ and $W$ are doing.

The readout probabilities derived in the main text are recovered as conditional probabilities in this history state. Conditioning on the clock reading $\mathcal{T}=K$ selects the final circuit state $\ket{\psi_K}$ of Eq.~\eqref{eq:readoutState}. Within that final state, conditioning on the readout record in $W$ selects the corresponding schedule outcome, and measuring $S$ then gives its conditional final-zero probability. The probabilities of the two records in Eq.~\eqref{eq:ppm} are $p_+=P(W{=}0\,|\,\mathcal{T}{=}K)$ and $p_-=P(W{=}1\,|\,\mathcal{T}{=}K)$, and the conditional final-zero probabilities are
\begin{equation}
\begin{aligned}
P_0^{(N,+)}&=P\bigl(S{=}0\,\big|\,W{=}0,\ \mathcal{T}{=}K\bigr),\\
P_0^{(N,-)}&=P\bigl(S{=}0\,\big|\,W{=}1,\ \mathcal{T}{=}K\bigr),
\end{aligned}
\end{equation}
These expressions reproduce Eqs.~\eqref{eq:clockOutcomeGeneral}, \eqref{eq:antizeno} and \eqref{eq:plusbranch}, since conditioning on $\mathcal{T}=K$ yields $\ket{\psi_K}$. 

The clock register $\mathcal{T}$ labels the stages of the circuit in our mathematical description. In the experiment, the control electronics determine when each gate is applied. They provide the timing needed to run the circuit, so a separate quantum clock register is not needed.

\section{Robustness of Visibility}
\label{app:visibility}

In this protocol, the visibility is robust against several common noise sources. This appendix shows that dephasing during the sequence, detuning, thermal preparation errors, and imperfect intermediate recording gates leave the visibility unchanged, provided three conditions hold: the offset rotation remains in the transverse plane, no $Q$-dependent dynamics act during the sequence, and the final recording faithfully correlates $S$ with $M_N$. We will then consider a pure $Z$-rotation, which makes the visibility sensitive to the Zeno final-zero probability.

\subsection{Visibility from Transverse Offset Rotations}
Let $\hat X: \mathcal{H}_S \rightarrow \mathcal{H}_{SME}$ be the isometry representing the full recording sequence, including any noisy evolution and coupling to an environment $E$. Here, $\operatorname{Ran}\hat X$ means all the states that the recording sequence $\hat X$ can produce. If the final recording is an ideal CNOT into a fresh memory $M_N$ initialized in $\ket{0}$, every output state has support only on $\ket{0}_S\ket{0}_{M_N}$ and $\ket{1}_S\ket{1}_{M_N}$, so the two computational-basis readings agree. Then, acting with $\sigma_x^S$ or $\sigma_y^S$ flips $S$ without flipping $M_N$, so $\sigma_{x,y}^S\operatorname{Ran}\hat{X}\perp \operatorname{Ran}\hat{X}$. Therefore,
\begin{equation}
    \hat{X}^\dagger \sigma_x^S \hat{X}=\hat{X}^\dagger \sigma_y^S \hat{X}=0.
\end{equation}
For any offset rotation on the equatorial axis, 
\begin{align}
    U_\delta=\cos{\frac{\Omega \delta}{2}}\mathbb{I}-i \sin{\frac{\Omega \delta}{2}} \mathbf{n_\perp}\cdot \boldsymbol{\sigma}\\
    \hat{X}^\dagger U_\delta \hat{X}=\cos{\frac{\Omega \delta}{2}}\mathbb{I}.
\end{align}
Then, for $\ket{\psi_{SM}^\mathcal{A}}=U_\delta \hat{X} \ket{0}$ and $\ket{\psi_{SM}^\mathcal{B}}=\hat{X} U_\delta \ket{0}$, we get the visibility
\begin{align}
    \mathcal{V}=\left|\braket{\psi_{SM}^{\mathcal{B}}|\psi_{SM}^{\mathcal{A}}}\right|=\bra{0}U_\delta^\dagger \hat{X}^\dagger U_\delta \hat{X} \ket{0}
    =\cos^2{\frac{\Omega \delta}{2}}.
\end{align}

The visibility therefore does not require a perfect $\sigma_x$ rotation. All the noise sources listed above can be absorbed into $\hat{X}$ without changing $\mathcal{V}$, provided they act before the final recording. The parity relation  $\hat{X}\sigma_x=\sigma_x^S \Sigma_M \hat{X}$, however, may not survive the same noise.

The visibility is also immune to thermal preparation errors. Suppose the initial state is a mixture rather than a pure state $\rho_S^{\text{ideal}}=\ket{0}\bra{0}$, for instance
\begin{equation}
    \rho_S=(1-p)\ket{0}\bra{0}+p \ket{1}\bra{1}=\frac{1}{2}[\mathbb{I}+(1-2p)\sigma_z],
\end{equation}
where $p=1/{(1+e^{{\epsilon}/{k_B T}})}$ with $\epsilon$ being the energy gap between the two states. Then, 
\begin{align}
    \braket{\psi_{SM}^{\mathcal{B}}|\psi_{SM}^{\mathcal{A}}}&=\operatorname{Tr}(\rho_S U_\delta^\dagger \hat{X}^\dagger U_\delta \hat{X})\\ \nonumber
    &=\cos{\frac{\Omega \delta}{2}} \operatorname{Tr}(\rho_S U_\delta^\dagger)\\ \nonumber
    &=\cos^2{\frac{\Omega \delta}{2}} \operatorname{Tr}(\rho_S)+\frac{i}{2}\operatorname{Tr}(\rho_S \mathbf{n_\perp}\cdot \boldsymbol{\sigma}) \sin{\Omega \delta}\\ \nonumber
    &=\cos^2{\frac{\Omega \delta}{2}},
\end{align}
as $\operatorname{Tr}(\rho_S \sigma_{x,y})=0$. The visibility remains $\mathcal{V}=|\braket{\psi_{SM}^{\mathcal{B}}|\psi_{SM}^{\mathcal{A}}}|=\cos^2{\frac{\Omega \delta}{2}}$. This holds for any initial state that is diagonal in the computational basis. Off-diagonal terms (x or y polarization) would change the visibility. In a superconducting implementation, however, thermal initialization errors produce a population mixture in the qubit’s bare energy basis, typically the computational $Z$ basis, so this condition is naturally satisfied.

\subsection{Pure $Z$-Rotation}

Consider the drive Hamiltonian of Eq.~\eqref{eq:HS}, $H=\frac{\Omega}{2}\sigma_x$, and let $\ket{\phi}=\hat{X}\ket{0}_S \ket{0...0}_M$ be the output of the zero-offset reference sequence. We can decompose $\ket{\phi}$ according to the final computational state of $S$
\begin{equation}
    \ket{\phi}=\ket{\phi_0}+\ket{\phi_1},
\end{equation}
where
\begin{align}
    \ket{\phi_0}&=(\ket{0}\bra{0}_S\otimes \mathbb{I}_M)\ket{\phi}\\
    \ket{\phi_1}&=(\ket{1}\bra{1}_S\otimes \mathbb{I}_M)\ket{\phi}.
\end{align}
They are orthogonal and satisfy 
\begin{align}
    \langle \phi_0|\phi_0\rangle=P_0^{(N)}=\frac{1}{2}[1+C_N]\\
    \langle \phi_1|\phi_1\rangle=P_1^{(N)}=\frac{1}{2}[1-C_N]
\end{align}
where $C_N=\cos^N(\Omega T/N)$. Now consider replacing the transverse offset rotation $U_\delta$ with a pure $Z$-rotation $R_Z(\varphi)=e^{-i \varphi \sigma_z/2}$. For two branches 
\begin{align}
    \ket{\psi_{SM}^{\mathcal{A}}}&=R_Z(\varphi) \hat{X}\ket{0}_S \ket{0...0}_M\\ \nonumber
    &=e^{-i \varphi/2}\ket{\phi_0}+e^{i \varphi/2} \ket{\phi_1},
\end{align}
and
\begin{align}
    \ket{\psi_{SM}^{\mathcal{B}}}&=\hat{X} R_Z(\varphi)\ket{0}_S \ket{0...0}_M\\ \nonumber
    &= e^{-i\varphi/2} \ket{\phi}.
\end{align}
Then the visibility will be 
\begin{align}
    \mathcal{V}(N)&=|\braket{\psi_{SM}^{\mathcal{B}}|\psi_{SM}^{\mathcal{A}}}|\\ \nonumber
    &=|\bra{\phi}\left(e^{-i \varphi/2}\ket{\phi_0}+e^{i \varphi/2}  \ket{\phi_1}\right)|\\ \nonumber
    &=|P_0^{(N)}+e^{i \varphi}P_1^{(N)}|\\ \nonumber
    &=\sqrt{1-4P_0^{(N)}P_1^{(N)}\sin^2(\varphi/2)}\\ \nonumber
    &=\sqrt{\cos^2(\varphi/2)+\sin^2(\varphi/2)\cos^{2N}(\frac{\Omega T}{N})}.
\end{align}
In the Zeno limit $N\rightarrow \infty$, $\mathcal{V}(N)\rightarrow 1$. At $\varphi=\pi$, $\mathcal{V}(N,\varphi=\pi)=|2P_0^{(N)}-1|=|P_0^{(N)}-P_1^{(N)}|$, i.e., visibility dictates the imbalance between final populations. A $Z$-rotation therefore makes the visibility sensitive to the Zeno final-zero probability, in contrast with the transverse case where $\mathcal{V}$ is independent of $N$.

\bibliographystyle{apsrev4-2}
\bibliography{references}

@book{Pauli1980,
  author    = {Pauli, Wolfgang},
  title     = {General Principles of Quantum Mechanics},
  publisher = {Springer},
  address   = {Berlin},
  year      = {1980},
  note      = {Translated by P. Achuthan and K. Venkatesan}
}

@article{DeWitt1967,
  author  = {DeWitt, B. S.},
  title   = {Quantum theory of gravity. I. The canonical theory},
  journal = {Phys. Rev.},
  volume  = {160},
  pages   = {1113},
  year    = {1967}
}

@article{PageWootters1983,
  author  = {Page, D. N. and Wootters, W. K.},
  title   = {Evolution without evolution: Dynamics described by stationary observables},
  journal = {Phys. Rev. D},
  volume  = {27},
  pages   = {2885},
  year    = {1983}
}

@article{Feynman1986,
  author  = {Feynman, R. P.},
  title   = {Quantum mechanical computers},
  journal = {Found. Phys.},
  volume  = {16},
  pages   = {507},
  year    = {1986}
}

@article{Wootters1984,
  author  = {Wootters, W. K.},
  title   = {{``Time'' replaced by quantum correlations}},
  journal = {Int. J. Theor. Phys.},
  volume  = {23},
  pages   = {701},
  year    = {1984}
}

@article{GLM2015,
  author  = {Giovannetti, V. and Lloyd, S. and Maccone, L.},
  title   = {Quantum time},
  journal = {Phys. Rev. D},
  volume  = {92},
  pages   = {045033},
  year    = {2015}
}

@article{MarlettoVedral2017,
  author  = {Marletto, C. and Vedral, V.},
  title   = {Evolution without evolution and without ambiguities},
  journal = {Phys. Rev. D},
  volume  = {95},
  pages   = {043510},
  year    = {2017}
}

@article{LeonMaccone2017,
  author  = {Le{\'o}n, J. and Maccone, L.},
  title   = {The Pauli objection},
  journal = {Found. Phys.},
  volume  = {47},
  pages   = {1597},
  year    = {2017}
}

@article{SmithAhmadi2019,
  author  = {Smith, A. R. H. and Ahmadi, M.},
  title   = {Quantizing time: Interacting clocks and systems},
  journal = {Quantum},
  volume  = {3},
  pages   = {160},
  year    = {2019}
}

@article{HohnSmithLock2021,
  author  = {H{\"o}hn, P. A. and Smith, A. R. H. and Lock, M. P. E.},
  title   = {Trinity of relational quantum dynamics},
  journal = {Phys. Rev. D},
  volume  = {104},
  pages   = {066001},
  year    = {2021}
}

@article{Moreva2014,
  author  = {Moreva, E. and Brida, G. and Gramegna, M. and Giovannetti, V. and Maccone, L. and Genovese, M.},
  title   = {Time from quantum entanglement: An experimental illustration},
  journal = {Phys. Rev. A},
  volume  = {89},
  pages   = {052122},
  year    = {2014}
}

@article{Moreva2017,
  author  = {Moreva, E. and Gramegna, M. and Brida, G. and Maccone, L. and Genovese, M.},
  title   = {Quantum time: Experimental multitime correlations},
  journal = {Phys. Rev. D},
  volume  = {96},
  pages   = {102005},
  year    = {2017}
}

@article{Baumann2021,
  author  = {Baumann, V. and Del Santo, F. and Smith, A. R. H. and Giacomini, F. and Castro-Ruiz, E. and Brukner, {\v C}.},
  title   = {Generalized probability rules from a timeless formulation of Wigner's friend scenarios},
  journal = {Quantum},
  volume  = {5},
  pages   = {524},
  year    = {2021}
}

@misc{PintoCouto2026,
  author        = {Pinto Couto, M. and Maccone, L. and Catani, L. and Roncallo, S.},
  title         = {First-click time measurements},
  year          = {2026},
  eprint        = {2603.28623},
  archivePrefix = {arXiv},
  primaryClass  = {quant-ph}
}

@article{Oreshkov2012,
  author  = {Oreshkov, O. and Costa, F. and Brukner, {\v C}.},
  title   = {Quantum correlations with no causal order},
  journal = {Nat. Commun.},
  volume  = {3},
  pages   = {1092},
  year    = {2012}
}

@article{Chiribella2013,
  author  = {Chiribella, G. and D'Ariano, G. M. and Perinotti, P. and Valiron, B.},
  title   = {Quantum computations without definite causal structure},
  journal = {Phys. Rev. A},
  volume  = {88},
  pages   = {022318},
  year    = {2013}
}

@article{Procopio2015,
  author  = {Procopio, L. M. and Moqanaki, A. and Ara{\'u}jo, M. and Costa, F. and Alonso Calafell, I. and Dowd, E. G. and Hamel, D. R. and Rozema, L. A. and Brukner, {\v C}. and Walther, P.},
  title   = {Experimental superposition of orders of quantum gates},
  journal = {Nat. Commun.},
  volume  = {6},
  pages   = {7913},
  year    = {2015}
}

@article{Henderson2020,
  author  = {Henderson, L. J. and Belenchia, A. and Castro-Ruiz, E. and Budroni, C. and Zych, M. and Brukner, {\v C}. and Mann, R. B.},
  title   = {Quantum temporal superposition: The case of quantum field theory},
  journal = {Phys. Rev. Lett.},
  volume  = {125},
  pages   = {131602},
  year    = {2020}
}

@article{IonicioiuTerno2011,
  author  = {Ionicioiu, R. and Terno, D. R.},
  title   = {Proposal for a quantum delayed-choice experiment},
  journal = {Phys. Rev. Lett.},
  volume  = {107},
  pages   = {230406},
  year    = {2011}
}

@article{Stone1930,
  author  = {Stone, M. H.},
  title   = {Linear transformations in Hilbert space. III. Operational methods and group theory},
  journal = {Proc. Natl. Acad. Sci. USA},
  volume  = {16},
  pages   = {172},
  year    = {1930}
}

@article{MisraSudarshan1977,
  author  = {Misra, B. and Sudarshan, E. C. G.},
  title   = {The Zeno's paradox in quantum theory},
  journal = {J. Math. Phys.},
  volume  = {18},
  pages   = {756},
  year    = {1977}
}

@article{Itano1990,
  author  = {Itano, W. M. and Heinzen, D. J. and Bollinger, J. J. and Wineland, D. J.},
  title   = {Quantum Zeno effect},
  journal = {Phys. Rev. A},
  volume  = {41},
  pages   = {2295},
  year    = {1990}
}

@article{Diaz2025,
  author  = {Diaz, N. L. and Braccia, P. and Larocca, M. and Matera, J. M. and Rossignoli, R. and Cerezo, M.},
  title   = {Parallel-in-time quantum simulation via Page and Wootters quantum time},
  journal = {Phys. Rev. Research},
  volume  = {7},
  pages   = {033294},
  year    = {2025}
}

@article{FacchiPascazio2008,
  author  = {Facchi, P. and Pascazio, S.},
  title   = {Quantum Zeno dynamics: mathematical and physical aspects},
  journal = {J. Phys. A: Math. Theor.},
  volume  = {41},
  pages   = {493001},
  year    = {2008},
  doi     = {10.1088/1751-8113/41/49/493001}
}

@article{Slichter2016,
  author  = {Slichter, D. H. and M{\"u}ller, C. and Vijay, R. and Weber, S. J. and Blais, A. and Siddiqi, I.},
  title   = {Quantum Zeno effect in the strong measurement regime of circuit quantum electrodynamics},
  journal = {New J. Phys.},
  volume  = {18},
  pages   = {053031},
  year    = {2016},
  doi     = {10.1088/1367-2630/18/5/053031}
}

@article{Harrington2017,
  author  = {Harrington, P. M. and Monroe, J. T. and Murch, K. W.},
  title   = {Quantum Zeno Effects from Measurement Controlled Qubit-Bath Interactions},
  journal = {Phys. Rev. Lett.},
  volume  = {118},
  pages   = {240401},
  year    = {2017},
  doi     = {10.1103/PhysRevLett.118.240401}
}

@article{EgusquizaGaray2003,
  author  = {Egusquiza, I. L. and Garay, L. J.},
  title   = {Real clocks and the Zeno effect},
  journal = {Phys. Rev. A},
  volume  = {68},
  pages   = {022104},
  year    = {2003},
  doi     = {10.1103/PhysRevA.68.022104}
}

@article{CastroRuiz2020,
  author  = {Castro-Ruiz, E. and Giacomini, F. and Belenchia, A. and Brukner, {\v C}.},
  title   = {Quantum clocks and the temporal localisability of events in the presence of gravitating quantum systems},
  journal = {Nat. Commun.},
  volume  = {11},
  pages   = {2672},
  year    = {2020},
  doi     = {10.1038/s41467-020-16013-1}
}

@article{Chiribella2012,
  author  = {Chiribella, G.},
  title   = {Perfect discrimination of no-signalling channels via quantum superposition of causal structures},
  journal = {Phys. Rev. A},
  volume  = {86},
  pages   = {040301(R)},
  year    = {2012},
  doi     = {10.1103/PhysRevA.86.040301}
}

@article{Goswami2018,
  author  = {Goswami, K. and Giarmatzi, C. and Kewming, M. and Costa, F. and Branciard, C. and Romero, J. and White, A. G.},
  title   = {Indefinite Causal Order in a Quantum Switch},
  journal = {Phys. Rev. Lett.},
  volume  = {121},
  pages   = {090503},
  year    = {2018},
  doi     = {10.1103/PhysRevLett.121.090503}
}

@article{Zych2019,
  author  = {Zych, M. and Costa, F. and Pikovski, I. and Brukner, {\v C}.},
  title   = {Bell's theorem for temporal order},
  journal = {Nat. Commun.},
  volume  = {10},
  pages   = {3772},
  year    = {2019},
  doi     = {10.1038/s41467-019-11579-x}
}

@article{Lin2022,
  author  = {Lin, J.-D. and Huang, C.-Y. and Lambert, N. and Chen, G.-Y. and Nori, F. and Chen, Y.-N.},
  title   = {Space-time dual quantum Zeno effect: Interferometric engineering of open quantum system dynamics},
  journal = {Phys. Rev. Research},
  volume  = {4},
  pages   = {033143},
  year    = {2022},
  doi     = {10.1103/PhysRevResearch.4.033143}
}

@article{Oreshkov2019,
  author  = {Oreshkov, O.},
  title   = {Time-delocalized quantum subsystems and operations: on the existence of processes with indefinite causal structure in quantum mechanics},
  journal = {Quantum},
  volume  = {3},
  pages   = {206},
  year    = {2019},
  doi     = {10.22331/q-2019-12-02-206}
}

@book{Kitaev2002,
  author    = {Kitaev, A. Yu. and Shen, A. H. and Vyalyi, M. N.},
  title     = {Classical and Quantum Computation},
  series    = {Graduate Studies in Mathematics},
  volume    = {47},
  publisher = {American Mathematical Society},
  address   = {Providence, RI},
  year      = {2002}
}

@misc{SeidelParityNotebook2026,
  author       = {Seidel, Olivia and Sussman, Sara},
  title        = {{Memory parity with relaxation, dephasing, and readout errors}},
  year         = {2026},
  howpublished = {{Jupyter} notebook and supporting data, \url{https://github.com/OliviaSeidel/superposed_zeno_schedule/tree/4764018}},
  note         = {File \texttt{paper\_analysis/memory\_parity\_robustness.ipynb}, commit 4764018, version of 11 November 2026}
}

@misc{SussmanSeidel2026ZenoFeasibility,
  author       = {Seidel, Olivia and Sussman, Sara},
  title        = {{Zeno-clock circuit feasibility on Nighthawk and Heron}},
  year         = {2026},
  howpublished = {{Jupyter} notebook, \url{https://github.com/OliviaSeidel/superposed_zeno_schedule/tree/4764018}},
  note         = {File \texttt{zeno\_feasibility\_heron\_nighthawk.ipynb}, commit 4764018}
}

@article{Abbott2020,
  author  = {Abbott, Alastair A. and Wechs, Julian and Horsman, Dominic and Mhalla, Mehdi and Branciard, Cyril},
  title   = {Communication through coherent control of quantum channels},
  journal = {Quantum},
  volume  = {4},
  pages   = {333},
  year    = {2020},
  doi     = {10.22331/q-2020-09-24-333}
}

\end{document}